\documentclass[twocolumn]{aastex63}
\usepackage{amsmath}

\begin{document}

\title{Hydrodynamics and Nucleosynthesis of Jet-Driven Supernovae I:
Parameter Study of the Dependence on Jet Energetics}

\shortauthors{Leung, Nomoto and Suzuki}
\shorttitle{Collapsar Hydro and Nucleosynthesis}

\author[0000-0002-4972-3803]{Shing-Chi Leung}

\affiliation{Department of Mathematics and Physics, SUNY Polytechnic Institute, 100 Seymour Road, Utica, NY 13502, USA}

\affiliation{TAPIR, Mailcode 350-17, California Institute of Technology, Pasadena, CA 91125, USA}

\author[0000-0001-9553-0685]{Ken'ichi Nomoto}

\affiliation{Kavli Institute for the Physics and 
Mathematics of the Universe (WPI),The University 
of Tokyo Institutes for Advanced Study, The 
University of Tokyo, Kashiwa, Chiba 277-8583, Japan}

\author{Tomoharu Suzuki}
\affiliation{School of General Education, Chubu University, 1200 Matsumoto-cho, Kasugai, Aichi 487-8501, Japan}

\correspondingauthor{Shing-Chi Leung}
\email{leungs@sunypoly.edu}

\date{\today}
\received{Jul 13 2022}
\revised{Jan 17 2023}
\accepted{Feb 20 2023}
\published{May 9 2023}

\begin{abstract}

Rotating massive stars with initial progenitor masses $M_{\rm prog} \sim$ 25 $M_{\odot}$ -- $\sim$140 $M_{\odot}$ can leave rapidly rotating black holes to become collapsars. The black holes and the surrounding accretion disks may develop powerful jets by magneto-hydrodynamics instabilities. The propagation of the jet in the stellar envelope 
provides the necessary shock heating for triggering nucleosynthesis unseen in canonical core-collapse supernovae. Yet, the energy budget of the jet and its effects on the final chemical abundance pattern are unclear. In this exploratory work, we present a survey on the parameter dependence of collapsar nucleosynthesis on jet energetics. We use the zero-metallicity star with $M_{\rm prog} \sim$ 40 $M_{\odot}$ as the progenitor. The parameters include the jet duration, its energy deposition rate, deposited energy, and the opening angle.
We examine the correlations of following observables: (1) the ejecta and remnant masses, (2) the energy deposition efficiency, (3) the $^{56}$Ni production and its correlation with the ejecta velocity, deposited energy, and the ejected mass, (4) the Sc-Ti-V correlation as observed in metal-poor stars, and
(5) the [Zn/Fe] ratio as observed in some metal-poor stars. 
We also provide the chemical abundance table of these explosion models for the use of the galactic chemical evolution and stellar archaeology.

\end{abstract}





\section{Introduction}

\subsection{Jet-induced Explosion and Gamma-Ray Burst (GRB)}

The association of gamma-ray bursts (GRBs) with massive star
explosions has stimulated vast interest in the last decades \citep[see a review, e.g.,][]{Tsuruta2018}.
Several well observed examples
include GRB980425 \citep[SN 1998bw: ][]{Galama1998}, and
GRB030329 \citep[SN 2003dh: ][]{Hjorth2003,Stanek2003}.
They show a much higher peak luminosity than canonical core-collapse supernovae
such as SN 1987A occurred in the Large Megallanic Cloud \citep[see a review, e.g.,][]{Nomoto1994b}.
These events hint at an explosion energy typically ten times above ordinary supernovae, named as hypernovae \citep{Iwamoto1998}. 
The explosion of a bare CO core (Type Ic supernova -- SN Ic) can resemble with many features observed in these events \citep[see, e.g.,][]{Woosley1993b}. The removal of H- and He-envelopes could be a result from tidal interaction from its companion in a binary system \citep{Nomoto1994a} or by efficient stellar wind mass loss.
The bifurcation of high mass star explosions into hypernovae and faint-SN branches \citep{Nomoto2010,Moriya2010} indicates the presence of an inner energy source after the core-collapse event.

The progenitor CO core is likely formed from stars with initial masses of $M_{\rm prog} \sim$ 25 -- 140 $ M_{\odot}$ \citep{Heger2002a}.
For $M_{\rm prog} \gtrsim 80$ $M_{\odot}$, the electron-positron pair instability induces significant pulsation and surface mass ejection before its final explosion 
\citep[e.g.,][]{Ohkubo2009,Takahashi2016,Woosley2017,Woosley2018,Woosley2019,Leung2019PPISN}.
The CO core formed in these stars is so compact 
that the bounce shock fails to disrupt the entire star and results in the black hole formation \citep{Sukhbold2016,Powell2021}. 
 
The jet forming accretion-disk is suggested to be closely related to metal-poor stars in view of nucleosynthesis and mixing \citep{Nomoto2010}. The suppressed mass loss in a lower metallicity star allows the star to maintain its angular momentum, so that
the remnant black hole becomes rapidly rotating. Furthermore, the angular momentum of the infalling matter supports the formation of an accretion disk. A minimum rotation is necessary for sustaining the accretion disk and launching the jet \citep{MacFadyen1999}.

While the jet is an essential component in this model, the exact formation mechanism remains unclear. It is a matter of debate whether the jet is driven by magneto-hydrodynamical (MHD) instabilities, by neutrinos, or by radiation. For the neutrino case \citep{Liu2017},
the extremely high temperature at the inner boundary of the accretion disk ($\sim$ 10 - 20 MeV) provides the necessary neutrino heat deposition as an energy source for the jet; however, the jet may not be strong enough to sustain until the stellar surface \citep{Wei2019}. 
For the radiation case, the explosion is similar to the MHD case but the photons may destroy most of the metal nuclei during its propagation to the surface \citep{Shibata2015}. The jet stability is also closely related to the jet geometry such as the opening angle \citep{Aloy2000,Zhang2003,Zhang2004}.

\subsection{Observational Hints from Stars}


Some low-metallicity stars show clear signs of collapsar explosion such as the indicative Zn production \citep{Maeda2003b, Aoki2014}.
The presence of stellar survey \citep[e.g., SAGA database; ][]{Suda2011} has largely extended the catalogue of stellar abundance, especially those with a 
low metallicity ($< 10^{-2}~Z_{\odot}$), which can be enriched by single or a few explosions \citep{Hartwig2019}. The abundance patterns of some carbon-enhanced metal-poor stars provide direct evidences on these asymmetric explosion models \citep{Tominaga2009}, e.g., HE 1327-2326 \citep{Ezzeddine2019}. The explosion morphology of ejected $^{44}$Ti and $^{56}$Ni can hint on the explosion history \citep{Magkotsios2010}.


\subsection{Motivation}

The stellar evolution  \citep[see, e.g,][]{Woosley1993a,Ohkubo2009,Nomoto2013,Woosley2015}
and nucleosynthesis \citep[see e.g.,][]{Woosley1995,Heger2002a,Tominaga2007a,Limongi2012,Nomoto2013,Umeda2017,Grimmett2018}
of (low metallicity) massive stars have been systematically studied. 
We notice that the jet propagation breaks the spherical symmetry assumed by these models. Multi-dimensional hydrodynamics simulations with nucleosynthesis are necessary to consistently trace the energy deposition of the jet and the associated nuclear reactions. \citep[see early works, e.g., ][]{Maeda2003a, Maeda2003b, Maeda2009, Couch2009, Nagataki2009}. The asymmetric energy deposition creates a high entropy flow within the jet-opening angle which synthesizes elements such as Ti, V, Cr and Zn. These elements are generally absent in a spherically symmetric model. Given the uncertainty of the jet energetics, it becomes interesting to investigate how the chemical composition can serve as an alternative constraint. 
We, therefore, carry out a parameter survey on the explosive nucleosynthesis of jet-driven supernovae. 
By comparing with some metal-poor stars, we explore the corresponding parameters of jet energetics that can reproduce the observed chemical abundances.

In Section \ref{sec:method} we present the numerical methods
used for modeling the jet-driven supernova. In Section \ref{sec:benchmark} we report our characteristic model, which aims at representing typical jet-driven supernovae,
and we examine its energetics and chemical abundance patterns.
In Section \ref{sec:survey} we present our 
parameter survey. We examine how the diversity of jet
in terms of the jet duration, energy deposition rate, deposited energy and jet open-angle.
In Section \ref{sec:nucleo} we examine the chemical abundance patterns and how they vary with each of the parameter explored. 
In Section \ref{sec:discussion} we compare our models
with those in the literature. Then we discuss in details
the possible observables of our explosion models, including
the remnant black hole mass, correlations among ejecta
mass, energy, velocity and $^{56}$Ni mass. After that
we further compare the Sc-Ti-V correlation and the
high [Zn/Fe] ratio as observed in metal poor stars reported recently in the 
literature. At last we give our conclusions.

\section{Numerical Method}
\label{sec:method}

In this section we briefly describe the methodology of our studies.
We solve the two-dimensional special relativistic Euler equations
in spherical coordinates, namely
\begin{equation}
\frac{\partial {\bf D}}{\partial t} + \nabla \cdot {\bf F} = {\bf S}, 
\end{equation}
where ${\bf D} = (\rho \Gamma, \rho \Gamma^2 h {\bf v}, \rho \Gamma^2 h - p - \rho \Gamma)^T$,
${\bf F} = (\rho \Gamma {\bf v}, \rho \Gamma^2 h {\bf v} {\bf v}, \rho \Gamma^2 h {\bf v} - D {\bf v})^T$ ,
${\bf S} = (0, \rho \nabla \Phi, \rho {\bf v} \cdot \nabla \Phi)^T$. 
Here $\rho, p, \epsilon, {\bf v}$ are the density, pressure,
specific internal energy and the velocity of the fluid. 
$h = 1 + p / \rho + \epsilon$ is the specific enthalpy
of the fluid. $\Phi$ is the gravitational potential which satisfies 
$\nabla^2 \Phi = 4 \pi G \rho$. $\Gamma = 1 / \sqrt{1 - {\bf v}^2}$
is the Lorentz contraction factor. 

This hydrodynamics code
is an extension of our previous 2D hydrodynamics code which models
the explosion phase of Type Ia supernovae by solving the 
Newtonian Euler equations \citep{Leung2015a}. The code uses the 
fifth-order weighted essentially non-oscillatory scheme for 
spatial discretization \citep{Shu1999} and the five-step third-order 
non-strong stability preserving Runge-Kutta scheme for the 
time-discretization \citep{Wang2007}. The code has been validated to reproduce classical 1D and 2D
numerical tests \citep{Leung2015a}. The code 
has been used for Type Ia supernovae with various mechanisms \citep{Leung2015b, Leung2018Chand, Leung2018SubChand}, electron capture supernovae \citep{Zha2019ECSN, Leung2018ECSN} and accretion induced collapse \citep{Leung2019AIC, Zha2019AIC}.

To close the equations, we use the Helmholtz equation of
state \citep[EOS, ][]{Timmes1999Helm, Timmes2000}. The EOS includes components from electrons of arbitrary degeneracy and relativistic levels, nuclei as ideal gases, $e^-$--$e^+$ pairs and photons with the Planck distribution. This subroutine takes the matter density, temperature, mean atomic number $\bar{Z}$
and mean atomic mass $\bar{A}$ as inputs, and compute relevant
thermodynamics quantities. 
To describe the local chemical composition, we use the 
7-isotope network which contains $^{4}$He, $^{12}$C, 
$^{16}$O, $^{20}$Ne, $^{24}$Mg, $^{28}$Si, $^{56}$Ni \citep{Timmes2000Iso}.
$^1$H-envelope is not considered because it is very extended from the star and does not contribute to the jet dynamics. The isotopes are modeled as scalars and follow the same advection scheme. 

To compute nucleosynthesis, we use the tracer particle scheme \citep[e.g., ][]{Travaglio2004} which records the  density and temperature evolution along the trajectories of the fluid elements. The particles are passive that they do not affect the fluid motion and they only follow the underlying fluid motion. The thermodynamics histories are used for post-processing of the nuclear reactions by a much larger network. In this
work, we use the 495-isotope network which 
contains nuclear reactions from $^{1}$H to $^{91}$Tc \citep{Timmes1999}.

We use a resolution of 300$\times$60 in spherical coordinates for the $(r, \theta)$ plane with an exponentially
increasing grid size in the radial direction and a constant grid size of $\pi / 120$ in the 
angular direction. The boundaries of the angular
and the inner radial directions are assumed to be reflecting while the outer boundary of the radial direction is set to be outflow. 
The progenitor model is the $M_{\rm prog} = 40 M_{\odot}$ zero-metallicity star as computed in \cite{Umeda2005,Tominaga2007b}. The simulations
are terminated when all tracer particles become sparse and
cold enough that no significant nuclear reaction can 
carry out and the ejecta forms homologous expansion to a 
good approximation. 

The jet is characterized by five parameters \citep{Tominaga2009}: (1) the total deposited energy $E_{{\rm dep}}$, (2) the 
energy deposition rate $\dot{E}_{{\rm dep}}$, (3) the Lorenz factor of the jet $\Gamma_{{\rm jet}}$, (4) the jet opening angle $\theta_{{\rm jet}}$ and 
(5) the thermal energy proportion $f_{th}$. We refer the interested readers to the derivation of the jet thermodynamics quantities in the original article. 
We assume that the jet enters the computational domain from the inner boundary within the jet opening angle. The jet is radiation dominated and it is absorbed by fluid elements along its path 
according to the local opacity. 
In the simulations, the jet effects are added separately by operator splitting, where the jet is assumed to propagate only radially outward. 
The propagation of the jet energy and its deposition is calculated in each step assuming a constant opacity. Due to the compactness, in most cases the jet energy is deposited at the innermost cell within the open angle. However, for a very extended jet duration, the innermost non-vacuum grid cell can largely recede and the deposited energy takes more than one time step to propagate before it is absorbed. Thus the energy density of the jet is modeled as an independent quantity, so that its propagation and absorption is consistently described. 

\section{Characteristic Model}
\label{sec:benchmark}

\subsection{Progenitor Model}

We use the presupernova model of the $M_{\rm prog} = 40 M_{\odot}$ zero-metallicity star as the initial model \citep{Tominaga2007b}. Prior to the start of the simulation, the Fe core is removed from the simulation and is assumed to have formed a compact object. In Table \ref{table:char_model} we tabulate the important progenitor parameters and setting of the jet energetics.
In this work all progenitor models are assumed to be spherically symmetric and non-rotating.
We further discuss the implications of our approach in Section \ref{sec:caveats}.

\begin{table}[]
    \centering
    \caption{Massive star progenitor model used by the characteristic model and the setting of the jet.}
    \begin{tabular}{c c c}
        \hline
         Model & Mass $(M_{\odot})$ & Radius (km)\\ \hline
         Progenitor star & 40 & $2.0 \times 10^7$ \\ 
         He star mass & 15 & $6.0 \times 10^5$ \\
         C-O core mass & 13.9 & $6.1 \times 10^4$ \\
         Si core mass & 3.81 & $9.4 \times 10^3$ \\
         \hline
         Jet parameter & Variable & Value \\
         \hline
         Deposited energy & $E_{\rm dep}$ & $1.5 \times 10^{52}$ erg \\
         Energy deposition rate & $\dot{E}_{\rm dep}$ & $1.2\times10^{53}$ erg s$^{-1}$ \\ 
         Jet deposition time & $t_{\rm jet}$ & 0.125 s \\
         Jet opening angle & $\theta_{\rm jet}$ & 15$^{\circ}$ \\
         \hline
    \end{tabular}
    
    \label{table:char_model}
\end{table}
\subsection{Hydrodynamics}

Here we present the hydrodynamical and thermodynamical history of the characteristic model to outline the important characters of the jet-powered explosion. 
In each simulation, we characterize the jet by the duration of the jet, $t_{\rm jet,0}$ and the energy deposition rate, $\dot{E}_{\rm dep,0}$

\begin{figure*}
\centering
\includegraphics*[width=8.5cm]{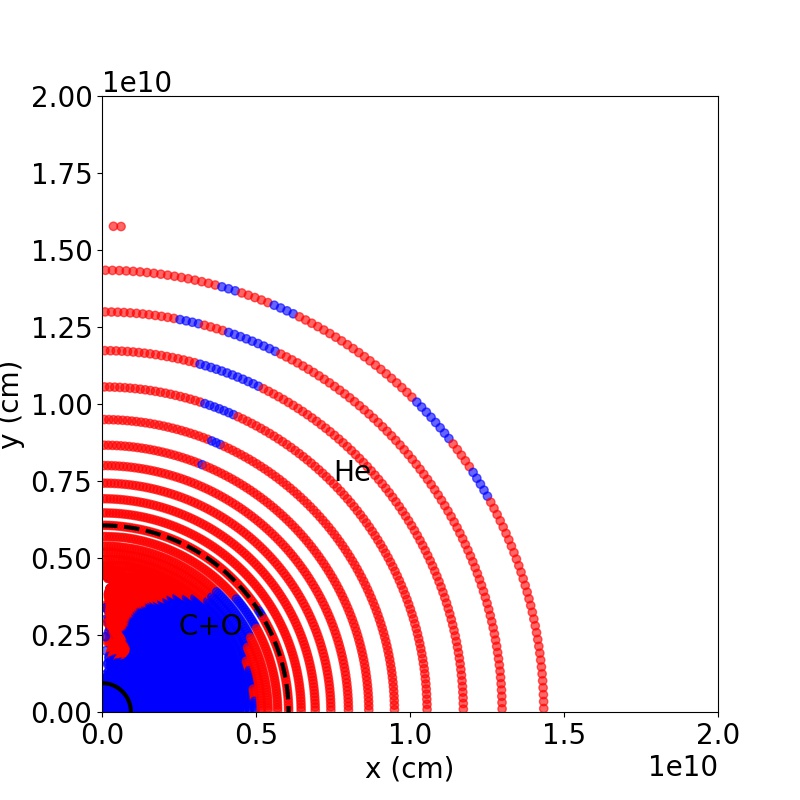}
\includegraphics*[width=8.5cm]{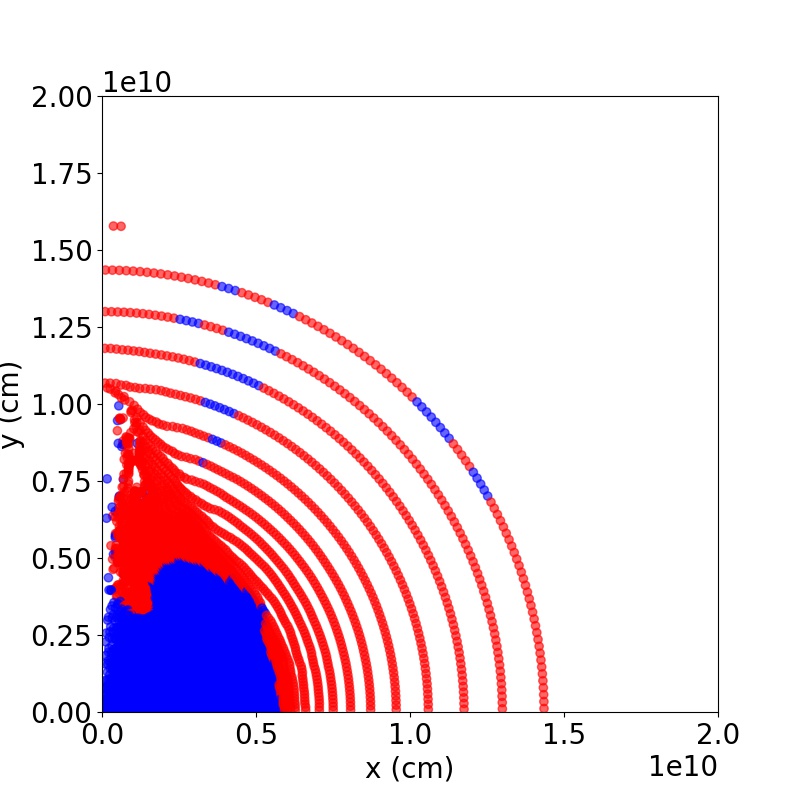}
\includegraphics*[width=8.5cm]{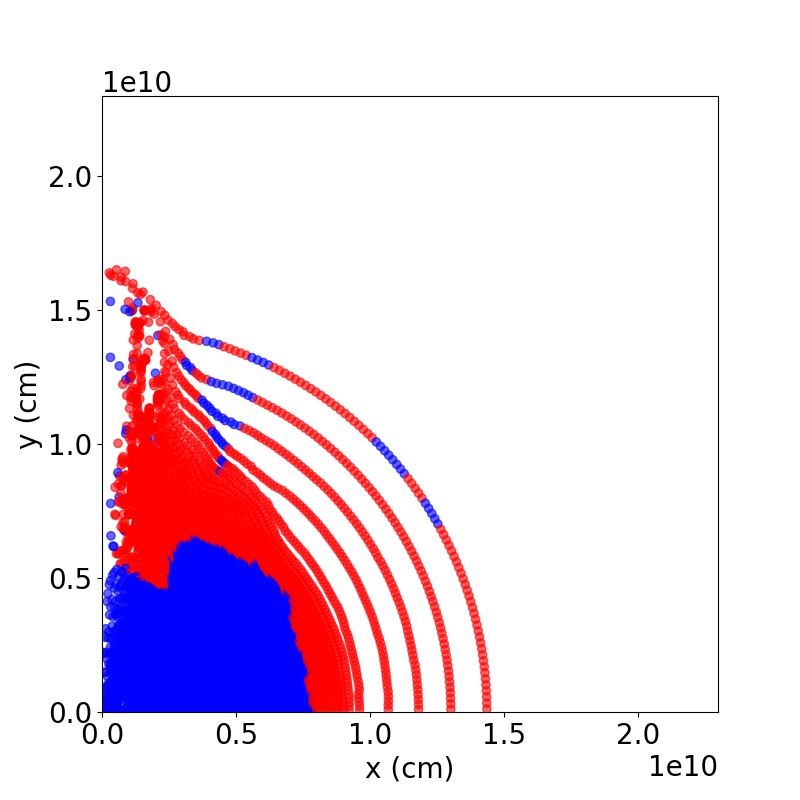}
\includegraphics*[width=8.5cm]{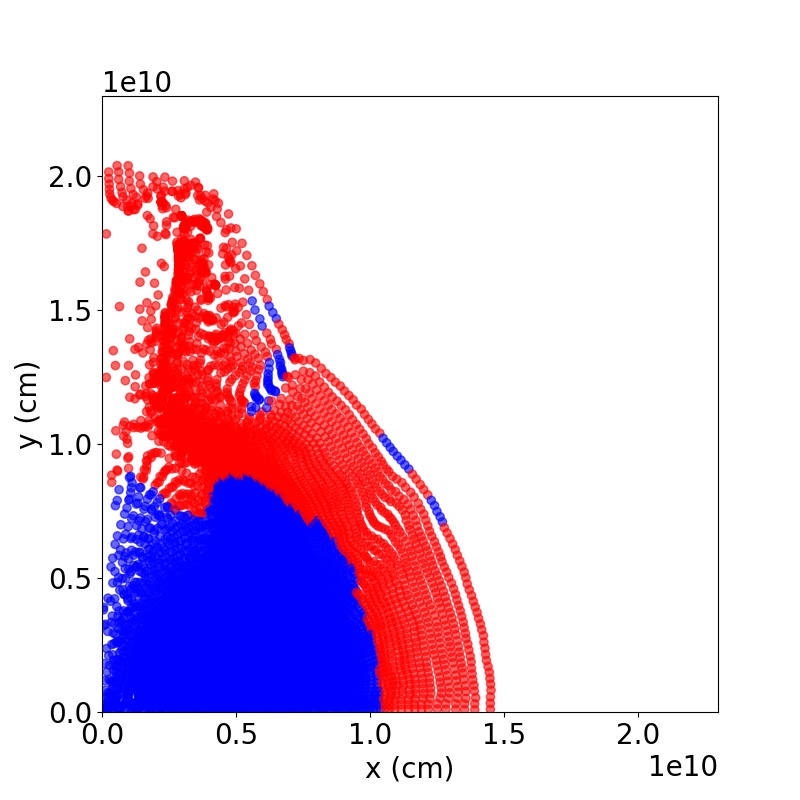}
\caption{The tracer particle distributions from 3.75 to 15 s 
at an interval of 3.75 s. The red and blue points
correspond to the tracer particles which can and cannot escape
by using their total energy as the criterion. In the top left panel, the solid (dashed) line stands for the outer boundary of the Si (C+O) core.}
\label{fig:tracer_benchmark}
\end{figure*}

In Figure \ref{fig:tracer_benchmark} we plot the tracer particles
of the characteristic model at 3.75, 7.50, 11.25 and 15.00 s.
The red and blue marks correspond to the tracers which 
can escape from the system and are bound by the system,
depending on their individual total energy $e = |v|^2/2 + \Phi(r)$,
where $v$ is the velocity and $\Phi$ is the local gravitational 
potential at the end of simulations. 
Some outer tracer particles are classified as bound because they have
a marginally negative energy. However, those tracers are likely to be ejected through shock compression as the density lowers. It takes $\sim$ 10 s for the shock to completely reach the envelope of the star (He-envelope extends to a much larger radius). The jet angle has increased to about $30^{\circ}$. At $t = 15$ s where the simulation ends, the jet accelerated particles have already broken out of the surface and reach as far as $2.0 \times 10^5$ km. The rapidly expanding flow swept away most matter along the cone shape with an opening angle $15^{\circ}$. The ejected matter gradually falls to the inner part of the simulation box, which will be later accreted. 

It is worthwhile to note that the final ejecta has an angular extension of $30^{\circ}$. The same phenomenon appears in other models besides the characteristic model. The expansion is due to the competition between thermal expansion along the angular direction and the radial propagation of the shock. A stronger shock implies a shorter travel time from the inner mass cut to the surface. At the same time, it also implies a stronger thermal expansion. The similarity of our models suggests that these effects approximately cancel each other.

\begin{figure}
\centering
\includegraphics*[width=8.5cm]{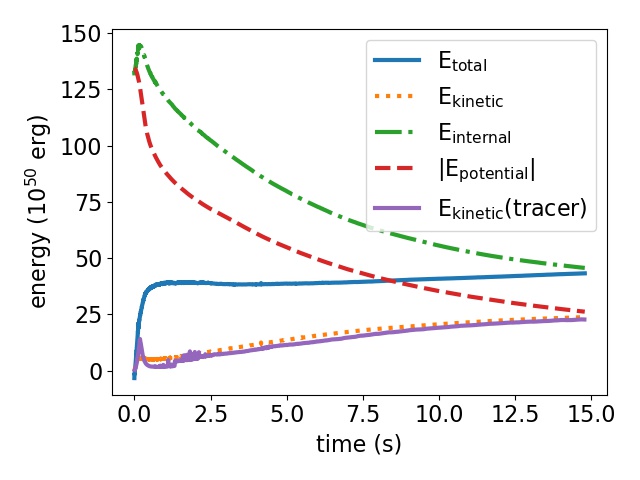}
\caption{The total energy, kinetic, internal and gravitational 
energy of the system in the characteristic model. The kinetic energy of the ejected tracer particles is also included for comparison. }
\label{fig:energy_benchmark}
\end{figure}

In Figure \ref{fig:energy_benchmark} we plot the total, kinetic, internal and gravitational energies of the characteristic model taken from the hydrodynamical simulation. Within the first one second, the jet has already finished injecting the energy to the system where the total energy increases to $\sim 4 \times 10^{51}$ erg. A small bump in the internal energy can also be seen. They show that the energy deposition from the jet creates a shock which provides significant shock heating by compression. After that, the internal energy drops, showing that the high velocity jet continues to lose its energy as work done to accelerate the outer matter in the star. Meanwhile, the kinetic and gravitational potential energies grow slowly. At $t = 10$ s, all energies reach their asymptotic values within $\sim 10\%$. Not all energy from the shock can be transferred to the ejecta because part of that is lost when the shock-heated fluid parcels expand and do work on the fluid elements along the angular direction. Also, near the bottom of the shock-heated fluid parcel, part of the matter falls back. They both dissipate the deposited energy. We also added the total kinetic energy of the tracers as a comparison. 



\begin{figure}
\centering
\includegraphics*[width=8.5cm]{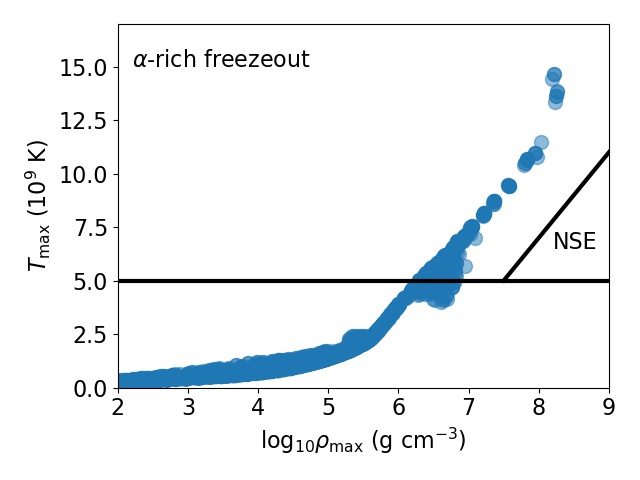}
\caption{The maximum temperature and density of the tracers 
particles experienced in the characteristic model. The lines separated the burning regime for $\alpha$-rich freezeout and nuclear statistical equilibrium (NSE).}
\label{fig:summary_benchmark}
\end{figure}

In Figure \ref{fig:summary_benchmark} we plot the thermodynamics history using the ejected
tracer particles. The distribution of particles 
shows two groups. In the low temperature branch, the tracers are not directly excited by the shock. They have a density from $10^2$--$10^5$ g cm$^{-3}$. 
They preserve the maximum temperature from the progenitor 
from $\sim 10^8$ to $2 \times 10^9$ K. For tracer particles in the high temperature branch, 
they are excited by the jet directly. They follow 
a steeper $\rho-T$ relation for the density range from $10^{5.5}$ to $10^{8.5}$ 
g cm$^{-3}$. Some ejected particles close to the mass cut initially
have a maximum temperature as high as $10 - 15 \times 10^9$ K. Between $\log_{10} \rho_{\rm max} = 6-7$, the cluster corresponds to the tracers indirectly excited by the jet. All tracers in the model do not enter nuclear statistical equilibrium (NSE). Instead they only achieve $\alpha$-rich freezeout\footnote{
The $\alpha$-rich freezeout \citep{Arnett1971, Woosley1973, Woosley1992} is the nuclear burning condition when the $\alpha$-particle abundance is high due to photodisintegration. The necessary condition for $\alpha$-rich freezeout to occur is the fast cooling time compared to the slowest He-burning. Therefore, not all the isotopes are equally accessible. This channel is important to explain the abundance of $^{57}$Fe, $^{59}$Co and $^{64}$Zn \citep{Thielemann1986}.
}. 



\begin{figure}
\centering
\includegraphics*[width=8.5cm]{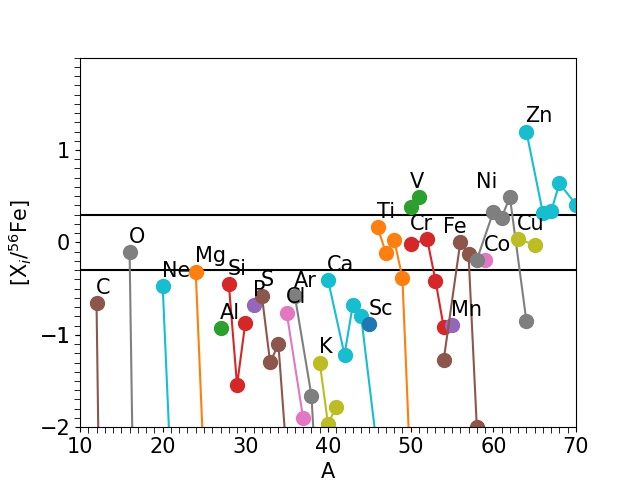}
\caption{$[X_i/^{56}$Fe]$= \log_{10} (X_i/^{56}$Fe)$/(X_i/^{56}$Fe$)_{\odot}$ for the characteristic model. The horizontal lines
correspond to two times and half of the solar value.}
\label{fig:final_benchmark}
\end{figure}

\subsection{Nucleosynthesis}

In Figure \ref{fig:final_benchmark} we plot the abundance pattern for 
the characteristic model. We calculate the post-process nucleosynthesis of the tracers until all major exothermic reactions cease. Then we wait for all short-lived radioactive isotopes to decay. We use two lines to show the abundance at two times and half of the solar values for comparison. Figure \ref{fig:final_benchmark} shows that: 

\noindent (1) For most of the lower mass elements from C to Ca, the abundance ratios of $\alpha$-chain elements to Fe are marginally compatible with the solar composition.

\noindent (2) Elements like Si, S and the nearby odd number elements are underproduced relative to Fe\footnote{In this article, discussion on elements or isotope abundances is all referring to the ratio to Fe}.

\noindent (3) On the contrary, elements from Ti onward are comparable with the solar composition.

\noindent (4) The high production of elements like V and Zn are consistent with the high entropy environment experienced in the shock heated matter.

\noindent (5) Mn, which is mostly produced by Type Ia supernovae, is also underproduced in our model.

\section{Parameter Survey of the Hydrodynamics}
\label{sec:survey}

\subsection{Model Description}

We first present the models studied in this work. For all models we name 
according to their configurations. For example, for 
Model S40-0250-4000-15, we use the 40 $M_{\odot}$ star model.
The energy deposition rate is 0.25 times of the characteristic 
model defined as $1.2 \times 10^{53}$ erg s$^{-1}$.
The energy deposition time is 4 times of the characteristic model
defined as 0.125 s. The jet opening angle is 15$^{\circ}$. 
We use this notation to describe all input physics we used to 
change the configuration of the jet. 
In Table \ref{table:models}
we tabulate the models studied in this work.
A total of 33 models are presented. 

\begin{table*}
\begin{center}
\caption{The models presented in this work. $M_{\rm prog}$ and $M_{{\rm ini}}$
are the zero-age main-sequence progenitor mass and the initial mass of the simulation
in units of $M_{\odot}$. $\dot{E}_{{\rm dep}}$ is the energy deposition rate
in units of $10^{51}$ erg s$^{-1}$. $t_{{\rm dep}}$ is the total energy
deposition time in unit of s. $\theta_{{\rm jet}}$ is the open-angle
of the jet in unit of degree. $E_{\rm dep}$ is the total energy
deposited by the jet in unit of $10^{51}$ erg. $M_{{\rm ej}}$ and $M(^{56}$Ni) are the ejecta
and included $^{56}$Ni mass in units of $M_{\odot}$.
$M$(Ti), $M$(V) $M$(Cr) and $M$(Zn) are the final stable masses of 
Ti, V, Cr and Zn in units of $(10^{-4} ~M_{\odot})$, $(10^{-4}~M_{\odot})$,
$(10^{-3} ~M_{\odot})$ and $(10^{-3} ~M_{\odot})$ respectively. The model with an asterisk is our characteristic model.
}
\begin{tabular}{l c c c c c c c c c c c c c }
\hline
Model & $M_{\rm prog}$ & $M_{{\rm ini}}$ & $\dot{E}_{{\rm dep}}$ & $t_{{\rm dep}}$ & $E_{{\rm dep}}$ & $\theta_{{\rm jet}}$ &   $M_{{\rm ej}}$ & $M(^{56}$Ni) & $M$(Ti) & $M$(V) & $M$(Cr) & $M$(Zn) \\ \hline
S40-1000-0125-15 & 40 & 15 & 120 & 0.015625 & 1.875 & 15 & 0.81  & 0.018 & 2.18 & 0.66 & 0.90 & $<0.01$ \\ 
S40-2000-0125-15 & 40 & 15 & 240 & 0.015625 & 3.750 & 15 & 0.57  & 0.037 & 2.75 & 0.33 & 0.66 & 0.01 \\ \hline
S40-0500-0250-15 & 40 & 15 &  60 & 0.031250 & 1.875 & 15 & 0.33  & 0.003 & 0.03 & 0.02 & 0.04 & $<0.01$ \\ 
S40-1000-0250-15 & 40 & 15 & 120 & 0.031250 & 3.750 & 15 & 0.25  & 0.013 & 0.38 & 0.25 & 0.50 & $<0.01$ \\ 
S40-2000-0250-15 & 40 & 15 & 240 & 0.031250 & 7.500 & 15 & 1.18  & 0.07 & 1.30 & 0.97 & 0.71 & 0.09 \\ \hline
S40-0250-0500-15 & 40 & 15 &  30 & 0.031250 & 1.875 & 15 & 0.50  & 0.005 & 0.77 & 0.22 & 0.44 & 0.02 \\ 
S40-0500-0500-15 & 40 & 15 &  60 & 0.062500 & 3.750 & 15 & 0.27  & 0.008 & 0.18 & 0.02 & 0.15 & $<0.01$ \\ 
S40-1000-0500-15 & 40 & 15 & 120 & 0.062500 & 7.500 & 15 & 1.17  & 0.07 & 2.02 & 1.56 & 1.08 & 0.03 \\ 
S40-2000-0500-15 & 40 & 15 & 240 & 0.062500 & 15.00 & 15 & 4.36  & 0.19 & 5.61 & 1.82 & 2.21 & 0.17 \\ 
S40-4000-0500-15 & 40 & 15 & 480 & 0.062500 & 30.00 & 15 & 9.56  & 0.57 & 14.1 & 2.38 & 9.05 & 4.98 \\ 
S40-8000-0500-15 & 40 & 15 & 960 & 0.062500 & 60.00 & 15 & 10.22 & 0.71 & 21.8 & 4.36 & 11.1 & 2.52 \\ \hline
S40-0125-1000-15 & 40 & 15 &  15 & 0.125000 & 1.875 & 15 & 0.04  & 0.003 & 0.11 & 0.03 & 0.04 & $<0.01$ \\ 
S40-0250-1000-15 & 40 & 15 &  30 & 0.125000 & 3.750 & 15 & 0.06  & 0.006 & 0.08 & 0.06 & 0.11 & $<0.01$ \\ 
S40-0500-1000-15 & 40 & 15 &  60 & 0.125000 & 7.500 & 15 & 1.34  & 0.076 & 2.31 & 1.78 & 1.23 & 0.03 \\ 
S40-1000-1000-7.5 & 40 & 15 & 120 & 0.125000 & 15.00 & 7.5 & 3.15 & 0.04 & 1.33 & 0.02 & 0.55 & 0.07 \\
S40-1000-1000-15* & 40 & 15 & 120 & 0.125000 & 15.00 & 15 & 4.13  & 0.30 & 7.31 & 2.57 & 4.17 & 4.81 \\ 
S40-1000-1000-30 & 40 & 15 & 120 & 0.125000 & 15.00 & 30 & 1.55  & 0.075 & 1.89 & 0.43 & 1.35 & 0.20 \\ 
S40-1000-1000-45 & 40 & 15 & 120 & 0.125000 & 15.00 & 45 & 0.90  & 0.005 & 0.01 & 0.13 & 0.12 & $<0.01$ \\ 
S40-2000-1000-15 & 40 & 15 & 240 & 0.250000 & 30.00 & 15 & 6.89  & 0.49 & 13.9 & 2.43 & 7.65 & 7.14 \\ 
S40-4000-1000-15 & 40 & 15 & 480 & 0.250000 & 60.00 & 15 & 10.21 & 0.65 & 13.9 & 2.50 & 10.3 & 5.48 \\ 
S40-8000-1000-15 & 40 & 15 & 960 & 0.250000 & 120.0 & 15 & 10.74 & 0.79 & 16.2 & 2.73 & 13.7 & 40.7 \\ \hline
S40-0125-2000-15 & 40 & 15 &  15 & 0.250000 & 3.750 & 15 & 0.65  & 0.055 & 6.51 & 1.02 & 1.95 & 0.02 \\ 
S40-0250-2000-15 & 40 & 15 &  30 & 0.250000 & 7.500 & 15 & 1.27  & 0.081 & 6.49 & 0.75 & 2.38 & 1.10 \\ 
S40-0500-2000-15 & 40 & 15 &  60 & 0.250000 & 15.00 & 15 & 2.39  & 0.13 & 9.72 & 2.38 & 1.93 & 3.75 \\
S40-1000-2000-15 & 40 & 15 & 120 & 0.250000 & 30.00 & 15 & 5.32  & 0.44 & 14.5 & 5.26 & 8.20 & 2.18 \\ 
S40-2000-2000-15 & 40 & 15 & 240 & 0.250000 & 60.00 & 15 & 8.10  & 0.59 & 15.3 & 2.74 & 9.35 & 4.83 \\ 
S40-4000-2000-15 & 40 & 15 & 480 & 0.250000 & 120.0 & 15 & 10.33 & 0.68 & 13.9 & 2.45 & 10.3 & 5.84 \\ \hline
S40-0250-4000-15 & 40 & 15 &  30 & 0.500000 & 15.00 & 15 & 1.62  & 0.15 & 4.08 & 1.99 & 2.95 & 4.52 \\ 
S40-0500-4000-15 & 40 & 15 &  60 & 0.500000 & 30.00 & 15 & 1.42  & 0.065 & 1.65 & 0.82 & 0.71 & 1.50 \\ 
S40-1000-4000-15 & 40 & 15 & 120 & 0.500000 & 60.00 & 15 & 5.59  & 0.46 & 1.13 & 3.40 & 7.91 & 4.28 \\ 
S40-2000-4000-15 & 40 & 15 & 240 & 0.500000 & 120.0 & 15 & 8.78  & 0.64 & 11.9 & 2.38 & 9.64 & 1.68 \\ \hline
S40-0500-8000-15 & 40 & 15 &  60 & 1.000000 & 60.00 & 15 & 3.02  & 0.21 & 6.31 & 2.40 & 7.96 & 5.22 \\ 
S40-1000-8000-15 & 40 & 15 & 120 & 1.000000 & 120.0 & 15 & 10.74 & 0.92 & 36.7 & 1.52 & 10.3 & 2.19 \\ \hline

\end{tabular}
\label{table:models}
\end{center}
\end{table*}

\subsection{Effects of Jet Duration}

\begin{figure}
\centering
\includegraphics*[width=8.5cm]{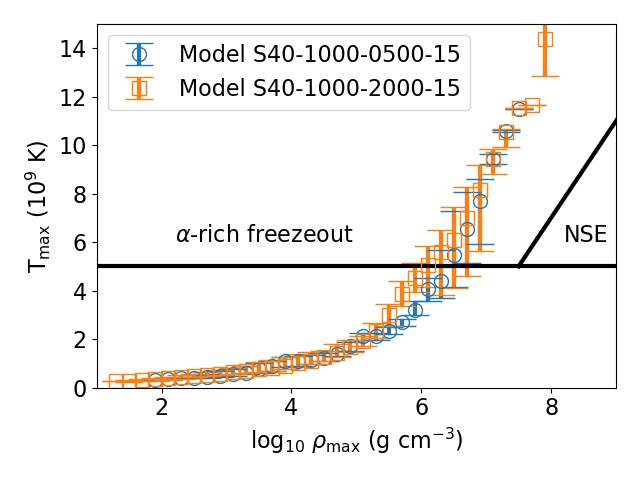}
\caption{
The $T_{\rm max}$ against $\rho_{\rm max}$ statistics for tracers from Models S40-1000-0500-15 ($t_{\rm jet} = 0.5 t_{\rm jet,0}$) and S40-1000-2000-15 ($t_{\rm jet} = 2 t_{\rm jet,0}$) . Both models use the same configurations of $\dot{E}_{\rm dep} = \dot{E}_{\rm dep,0}$ and $\theta_{\rm jet} = 15^{\circ}$. The error bar corresponds to the temperature range of tracers within the same density bins. Solid black lines are added to classify the burning categories. 
}
\label{fig:tracer_plot_jettime}
\end{figure}


We plot in Figure \ref{fig:tracer_plot_jettime} the thermodynamics
record for tracers for Models S40-1000-0500-15 and S40-1000-2000-15
by their maximum temperature and density, binned by their maximum density. We observe that 
the effects of the jet duration affects mostly the 
tracer particles near the core with a density $\geq 10^{5.5}$ g cm$^{-3}$.
The maximum temperature is $\sim 20 \%$ higher for the model
with a longer jet duration. It also triggers a wider spread
in the maximum temperature among particles. The
model with a higher jet duration also has tracer particles with 
higher $\rho_{{\rm max}}$. The longer energy 
deposition allows the shock to maintain its strength, which suppresses fall back into the central remnant. All tracers are burnt in either $\alpha$-rich freezeout or incomplete burning.

\subsection{Effects of Energy Deposition Rate}

\begin{figure}
\centering
\includegraphics*[width=8.5cm]{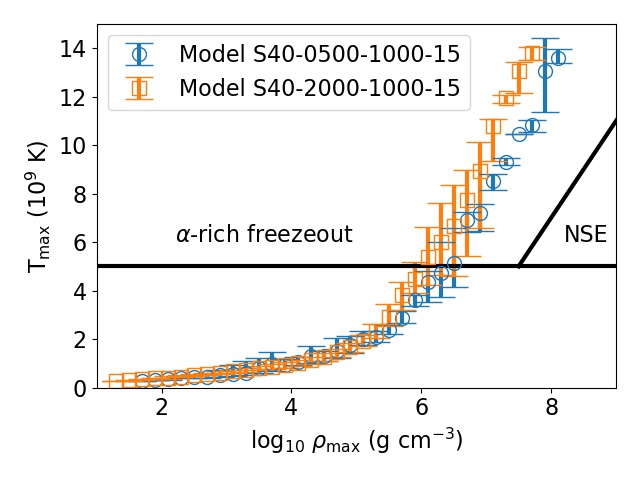}
\caption{
Same as Figure \ref{fig:tracer_plot_jettime}, but for tracers from Models S40-0500-1000-15($\dot{E}_{\rm dep} = 0.5 \dot{E}_{\rm dep,0}$) and S40-2000-1000-15 ($\dot{E}_{\rm dep} = 2 \dot{E}_{\rm dep,0}$). Both models use the same configurations of $t_{\rm jet} = t_{\rm jet,0}$ and $\theta_{\rm jet} = 15^{\circ}$. 
}
\label{fig:tracer_plot_dedep}
\end{figure}

%


In Figure \ref{fig:tracer_plot_dedep} we plot the thermodynamics
trajectories similar to Figure \ref{fig:tracer_plot_jettime}
but for Models S40-0500-1000-15 and S40-2000-1000-15. Again, for 
tracer particles with $\rho_{{\rm max}} > 10^{5.5}$ g cm$^{-3}$, derivation appears between the behaviors of the two models, driven by the differences in the adopted jet parameters. The model with a high energy deposition
again has a higher $T_{{\rm max}}$ and a wider spread. However, 
the $\rho_{{\rm max}}$ range of the two models is comparable.

\subsection{Effects of Jet Angle}

\begin{figure}
\centering
\includegraphics*[width=8.5cm]{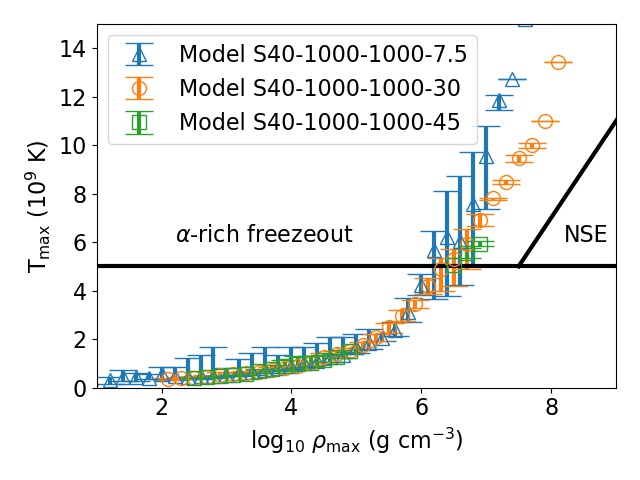}
\caption{Same as Figure \ref{fig:tracer_plot_jettime}, but for Models S40-1000-1000-7.5 ($\theta_{\rm jet} = 7.5^{\circ}$), S40-1000-1000-30 ($\theta_{\rm jet} = 30^{\circ}$) and S40-1000-1000-45 ($\theta_{\rm jet} = 45^{\circ}$). All models use the same configurations of $t_{\rm jet} = t_{\rm jet,0}$ and $\dot{E}_{\rm dep} = \dot{E}_{\rm dep,0}$.}
\label{fig:tracer_plot_angle}
\end{figure}


In Figure \ref{fig:tracer_plot_angle} we plot the statistics of the 
ejected tracer particles for Models S40-1000-1000-7.5, S40-1000-1000-30 and S40-1000-1000-45. 
The geometry of the shock plays an important
role in the thermodynamical evolution of the ejecta. At 
an angle above 30$^{\circ}$, the inner part of the ejecta has a temperature $\sim15 \times 10^9$ K
where some of the tracers in the Si-layer is also ejected.
When the 
jet angle increases, the large surface area makes the shock dissipate faster and the shock heating is less efficient. A lower temperature-maximum of $6 \times 10^9$ K
is recorded. The shock compression with the maximum density of $\sim 10^7$ g cm$^{-3}$ is also weaker than the previous case.
On the other hand, as $\theta$ decreases to 7.5 degrees, the more concentrated energy deposition leads to a stronger shock, with stronger heating and compression for a given $\rho_{\rm max}$. The distribution suggests that the jet energy has the largest effect when $\theta_{\rm jet} \approx 30^{\circ}$. This can be understood as the competition of two factors, the deposited mass and the shock strength. At a small jet angle, the initial shock is strong but the deposited mass is small. Thus, the effect of shock heating is limited. At a large jet angle, the shock is weak but more mass is affected.

\subsection{Isoenergetic Model}

\begin{figure}
\centering
\includegraphics*[width=8.5cm]{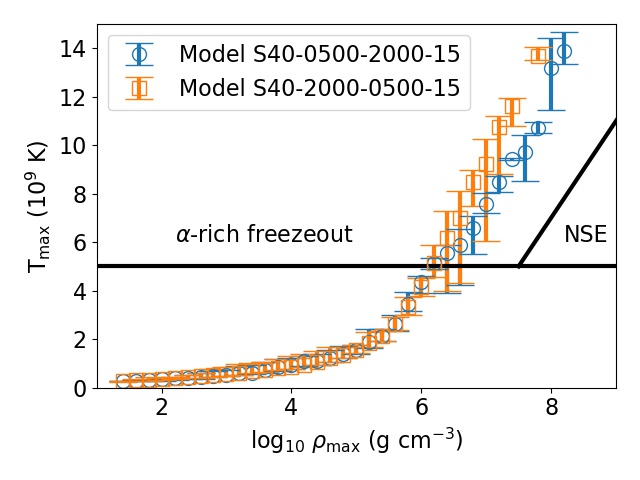}
\caption{Same as Figure \ref{fig:tracer_plot_jettime} but for Models S40-0500-2000-15 ($t_{\rm jet} = 2 t_{\rm jet,0}$ and $\dot{E}_{\rm dep} = 0.5 \dot{E}_{\rm dep,0}$) and S40-2000-0500-15 ($t_{\rm jet} = 0.5 t_{\rm jet,0}$ and $\dot{E}_{\rm dep} = 2 \dot{E}_{\rm dep,0}$). Both models use the same configurations of $\theta_{\rm jet} = 15^{\circ}$.}
\label{fig:tracer_plot_isoE}
\end{figure}


In Figure \ref{fig:tracer_plot_isoE} we plot the statistics of the 
ejected tracer particles for Models S40-0500-2000-15 and S40-2000-0500-15. The two models assume the same total deposition energy $E_{\rm jet}$ but with different $t_{\rm jet}$ and $\dot{E}_{\rm jet}$. Both models exhibit similarities in the statistic of tracers, especially at low $\rho_{\rm max}$. This suggests that the outer envelope is less sensitive to the jet characteristics. On the other hand, the differences in the high $\rho_{\rm max}$ demonstrates the sensitivity of the temperature range in the ejecta on $t_{\rm jet}$. A longer energy deposition helps inner tracers to escape from the star. 

It might look contradictory that S40-2000-050-15, which has a higher $T_{\rm max}$, has a lower ejected mass than S40-0500-2000-15. In fact, it is because the total ejected mass depends on two factors: the shock strength and its sustainability. With a shorter $t_{\rm jet}$, the early shock dissipates its energy and the matter in the inner region does not have sufficient energy input to maintain its expansion. On the other hand, a longer $t_{\rm jet}$ means a weaker shock. But the expansion is long enough that the matter from the inner core becomes unbound.



\subsection{Remarks}

In this section we have considered all three variables, $E_{\rm dep}$, $\dot{E}_{\rm dep}$ and $t_{\rm jet}$ as independent variables. While, by definition $E_{\rm dep} = \dot{E}_{\rm dep} t_{\rm dep}$, this means that these variables are not fully independent. Changing, for example $t_{\rm jet}$ while keeping $\dot{E}_{\rm dep}$ constant, still changes $E_{\rm dep}$. Note that all three variables are not yet well constrained by observational data. When we fix one of the variables, effectively we are observing the jet-dependence of the nucleosynthesis yield on one of the slices of the two-dimensional surface in the parameter space. Further constraints, such as a precise measurement of $E_{\rm dep}$ in real collapsars, will indicate which parameter ``slices'' presented in this section is necessary for the comparison.

\section{Nucleosynthesis Pattern}
\label{sec:nucleo}

\subsection{Effects of Jet Duration}

\begin{figure}
\centering
\includegraphics*[width=8.5cm]{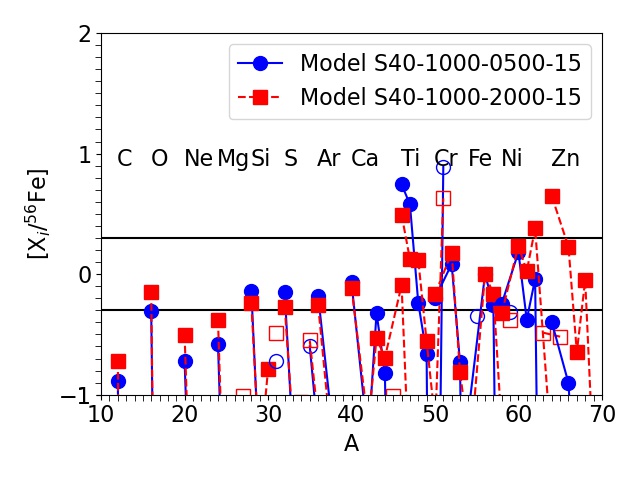}
\caption{$[X_i/^{56}$Fe] for Models S40-1000-0500-15 ($t_{\rm jet} = 0.5 t_{\rm jet,0}$) and S40-1000-2000-15 ($t_{\rm jet} = 2 t_{\rm jet,0}$) 
using their initial position. Both models use the same configurations of $\dot{E}_{\rm dep} = \dot{E}_{\rm dep,0}$ and $\theta_{\rm jet} = 15^{\circ}$. We assume all short-lived radioactive isotopes have decayed. The lines stand for 50 \% and 200 \% of the solar value. Odd number elements are plotted with unfilled symbols for contrast.}
\label{fig:final_jettime}
\end{figure}



We first examine how the chemical abundance patterns of the ejecta 
depend on the jet duration. We compare in Figure 
\ref{fig:final_jettime} the abundance pattern for two contrasting
models S40-1000-0500-15 ($t_{\rm jet} = 0.5 t_{\rm jet,0}$)
and S40-1000-2000-15 ($t_{\rm jet} = 2 t_{\rm jet,0}$). We remark that the two models also differ in also the total deposited energy, but the early shock, which we will show to largely change the Fe-group element synthesis, is identical in both models. We also remark that in this section, all elements are referred as the ratios to $^{56}$Fe, instead of the absolute values of the mass fraction.

The two models share similar abundance patterns for lighter
elements including O, Ne and Mg. Similar pattern can be observed for intermediate mass elements (IMEs) including Si, S, Ar and Ca. But there
is a minor enhancement in P and Sc for the model with a longer energy deposition. Iron-group elements (IGE) from Ti to Ni are also similar,
with over-production in $^{48}$Ti and $^{51}$V. Substantial 
differences are found for $^{55}$Mn and $^{64,66,67}$Zn.
The weaker explosion model has lower IGE abundances than the other by an order of magnitude. 




\subsection{Effects of Energy Deposition rate}

\begin{figure}
\centering
\includegraphics*[width=8.5cm]{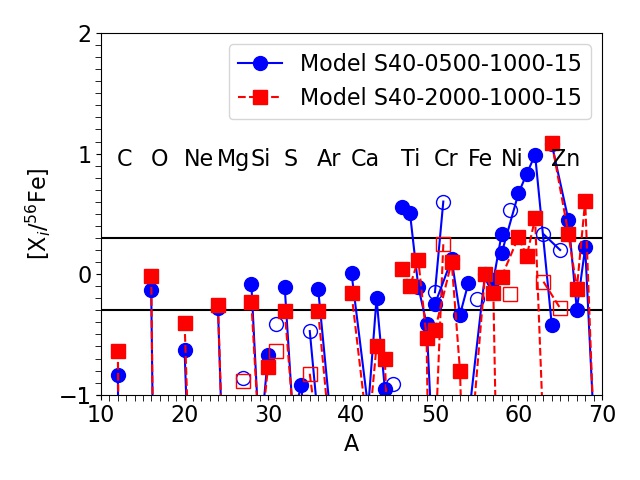}
\caption{Same as Figure \ref{fig:final_jettime} but for Models S40-0500-1000-15 ($\dot{E}_{\rm dep} = 0.5 \dot{E}_{\rm dep,0}$) 
and S40-2000-1000-15 ($\dot{E}_{\rm dep} = 2 \dot{E}_{\rm dep,0}$).
Both models use the same configurations of $t_{\rm jet} = t_{\rm jet,0}$
and $\theta_{\rm jet} = 15$ degree. }
\label{fig:final_dedep}
\end{figure}



In Figure \ref{fig:final_dedep} we plot the final abundance of Models S40-0500-1000-15 ($\dot{E}_{\rm dep} = 0.5 \dot{E}_{\rm dep,0}$)
and S40-2000-1000-15 ($\dot{E}_{\rm dep} = 2 \dot{E}_{\rm dep,0}$).
The two models differ by the energy deposition rate to 
be half and double of the characteristic model.
The model with a higher energy deposition rate ejects a larger fraction of near-surface material because of a higher energy deposited. As a result, a
higher abundance of C, O and Mg is found.
As the global $^{56}$Ni mass increases,
as well as more material from the deeper core
is ejected, the IME and the IGE
abundances are suppressed. On the contrary, for the 
model with a lower deposition rate, the much smaller
production of $^{56}$Fe (or $^{56}$Ni before decay) allows the formation
of peculiar abundance patterns. This includes a 
super-solar production of $^{46-47}$Ti,  $^{51}$V,
$^{59}$Co and $^{58-62}$Ni. The similar $^{64}$Zn production suggests both models have experienced a similar high entropy phase. 



\subsection{Effects of Jet injection Angle}

\begin{figure}
\centering
\includegraphics*[width=8.5cm]{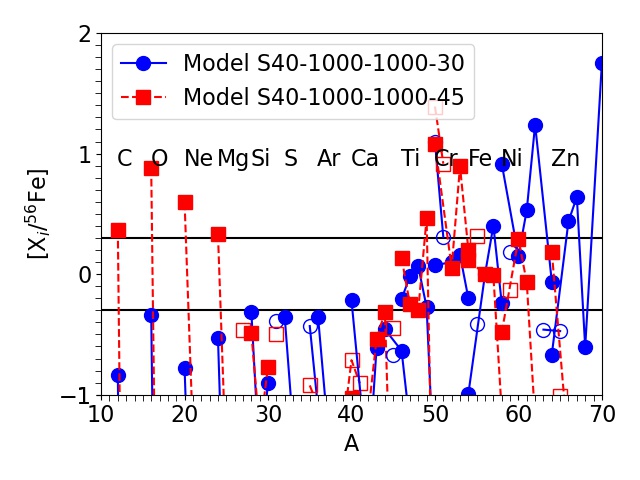}
\caption{Same as Figure \ref{fig:final_jettime} but for Models S40-1000-1000-30 ($\theta_{\rm jet} = 30$ deg) and S40-1000-1000-45 ($\theta_{\rm jet} = 45$ deg).
Both models use the same configurations of 
$\dot{E}_{\rm dep} = \dot{E}_{\rm dep,0}$ and $t_{\rm jet} = t_{\rm jet,0}$.
}
\label{fig:final_angle}
\end{figure}


In Figure \ref{fig:final_angle} we plot $[X_i/^{56}$Fe] for
Models S40-1000-1000-30 ($\theta_{\rm jet} = 30^{\circ}$) and S40-1000-1000-45 ($\theta_{\rm jet} = 45^{\circ}$) . The two models differ
from each other by the opening angle, which is two or three
times of the characteristic model. When the jet opening angle increases,
the more pronounced envelope ejection is reflected by the 
enhanced abundance of C, O, Ne and Mg. The IME production is suppressed. The ratios of Cr, Fe and Mn are enhanced due to the lower $^{56}$Fe production. But elements yielded in typical
alpha-rich freezeout, such as Ni and Zn, are not well produced. 



\subsection{Comparing Isoenergtic Model}

\begin{figure}
\centering
\includegraphics*[width=8.5cm]{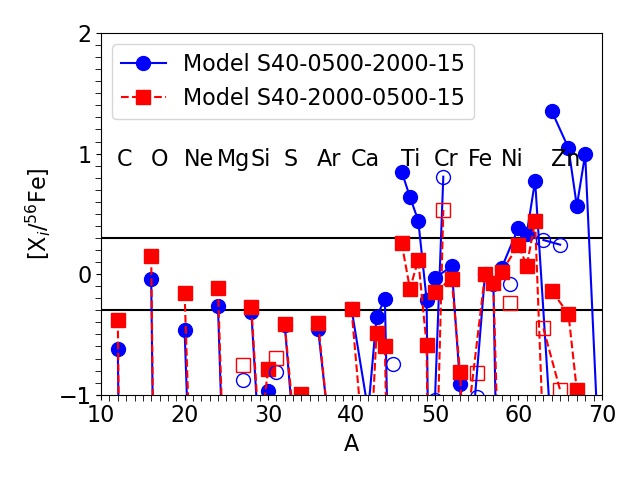}
\caption{Same as Figure \ref{fig:final_jettime} but for Models S40-0500-2000-30 ($\dot{E}_{\rm dep} = 0.5 \dot{E}_{\rm dep,0}$ and $t_{\rm jet} = 2 t_{\rm jet,0})$ and S40-2000-0500-15 ($\dot{E}_{\rm dep} = 2 \dot{E}_{\rm dep,0}$ and $t_{\rm jet} = 0.5 t_{\rm jet,0)}$.
Both models use the same configurations of 
$\theta_{\rm jet} = 15^{\circ}$.
}
\label{fig:final_isoE}
\end{figure}

In Figure \ref{fig:final_isoE} we plot $[X_i/^{56}$Fe] for the Models
S40-2000-0500-15 and S40-0500-2000-15. The two models differs
from each other by the jet duration and energy deposition rate while the total deposited energy is fixed.
The abundance patterns of the two models are similar. Elements from C to Ca and from
Fe to Ni show very 
good agreement with each other. In particular, the 
alpha-chain isotopes, such as $^{28}$Si, $^{32}$S, $^{36}$Ar
and $^{40}$Ca show almost complete overlap. 

Model S40-0500-2000-15
has a more pronounced Ti, V, Cu and Zn production. These are
the elements which require high entropy for the formation.
A long $t_{\rm jet}$ is important in synthesizing these elements.

\section{Discussion}
\label{sec:discussion}

In this section, we discuss the implications of our kinematics and nucleosynthesis results.
First, we compare our models with those in the literature. 
Then, we further extract the observables of our models 
to study their parameter dependence, and compare with observational
data. They include the jet-dependence of (1) the remnant or ejecta mass, 
(2) efficiency of energy deposition (3) $^{56}$Ni production, 
(4) $^{56}$Ni mass against ejecta velocity relation, 
(5) Sc-Ti-V relations, and (6) the [Zn/Fe] ratio. 

\subsection{Comparison with Models in the Literature}

There are not many works in the literature which extensively cover the effects of the jet on hydrodynamics and nucleosynthesis presented in 
this work. \cite{Tominaga2009} presents contrasting models which 
explicitly studies how the energetics of the jet affects
the associated nucleosynthesis. Here we compare the 
input physics used in their work and the numerical results.
In Table \ref{table:method} we compare the numerical algorithms used in their work and in this work.

\begin{table*}[]
    \centering
    \caption{Comparison of the collapsar model of this work and \cite{Tominaga2009}}
    \begin{tabular}{c c c}
    \hline
         & This work & \cite{Tominaga2009} \\ \hline
        Code & \cite{Leung2015b} & \cite{Donat1998} \\ 
        Spatial discretization & $5^{\rm nd}$ order \citep{Shu1999} & $2^{nd}$ order  \citep{Marquina1994}\\
        Temporal discretization & $3^{rd}$ order \citep{Wang2007} & $3^{\rm rd}$ order \citep{Aloy1999} \\
        Equation of state & Helmholtz \citep{Timmes2000} & ideal gas + $e^-e^+$-pair \citep{Freiburghaus1999} \\
        Gravity solver & Gaussian relaxation & Spherical harmonics \\
        Post-processing & 495-isotope network & 280-isotope network \\
        Progenitor & 40 $M_{\odot}$ zero metallicity star & 40 $M_{\odot}$ zero metallicity star \\ \hline
    \end{tabular}
    
    \label{table:method}
\end{table*}



For the progenitor, we use the same 40 $M_{\odot}$
zero-metallicity star at the onset of collapse as the progenitor.
While the exact grid and hydrodynamical schemes are different, we choose a compatible choice of mass-cut and jet energetics. We make reference on their Model A 
as our Model S40-1000-1000-15.
They observe the remnant of mass 9.1 $M_{\odot}$
while ours is $\sim 10.9 ~M_{\odot}$. The larger 
remnant mass is because we observe the very outer envelope could include very outer He-envelope which is extended and low in binding energy. 
The prescription ensures that net energy is deposited to outgoing matter. 
This adds an extra $\sim 1~M_{\odot}$ to the ejecta mass. Notice that this does not change the nucleosynthetic pattern because the density and temperature in the He-envelope is too low for significant reactions to occur.

Figure \ref{fig:tracer_benchmark} shows a very similar 
distribution of ejecta and remnant structure as their Figure 4a. 
The whole He layer and C+O layer are ejected after the explosion.
A disk shape structure which is bound concentrates near the innermost Si-layer and 
C+O layer. 

For nucleosynthesis, our model predicts a higher Fe production, which leads to a lower C, O and Ne ratios
and the global abundance pattern as the abundances are taken as the ratio to $^{56}$Fe.
Both our and their works show an underproduction of Sc and Mn. 
Their model shows a flat distribution of Ti, V and Cr, while V in our model is slightly over-produced and is higher than Ti and Cr. At last, for IGEs, our model shows a higher stable Ni, Cu and Zn, suggesting more contribution from inner ejecta.

Globally, our model consists of more tracers with complete burning, especially alpha-rich freezeout elements.
The difference of nucleosynthesis is closely related
to the EOS and the nuclear reaction network available. The exact temperature of the shock front can be sensitive to the equation of states and the numerical shock capturing scheme. The numerical dissipation may also affect how the shock heating behaves in the outer layer of the stellar envelope.

\subsection{Remnant Black Hole Mass}

\begin{figure}
\centering
\includegraphics*[width=8.5cm]{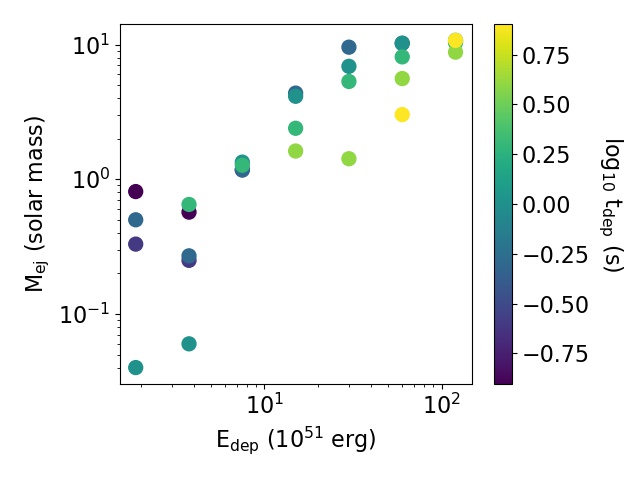}
\caption{The ejecta mass against deposited energy from the 
models presented in this work.}
\label{fig:Mej_dist}
\end{figure}

Recent observations of gravitational wave signals measured
by advanced-LIGO and VIRGO demonstrated the existence of black hole-black hole
mergers and neutron star-neutron star mergers \citep{Abbott2018}.
The third observation run \citep{Abbott2021a} has discovered a wide distribution of black hole mass and spins. The black hole statistics \citep{Abbott2021b} could be directly related to the mass ejection process of this class of supernovae. 

In Figure \ref{fig:Mej_dist} we plot the ejecta mass distribution
as a function of the deposited energy $E_{\rm dep}$. As $E_{\rm dep}$ increases, the ejected mass increases significantly and approaches an asymptotic value. Almost the entire
envelope is accreted in the low energy limit. In the high energy limit, about two-third
of the envelope is ejected. 
This corresponds to the black hole mass range from 5 to 
15 $M_{\odot}$. The asymptotic limit exists because the Si- and C+O-core outside the jet cone cannot be directly excited by this mechanism. They always fall back and remain bound.
Below $\sim 2 \times 10^{51}$ erg, the jet
fails to eject any observable mass.

Early modeling of the light curve and spectra from SN 1998bw \citep[e.g.,][]{Maeda2002} suggested an explosion energy of $\sim 10^{52}$ erg, and an ejecta mass 
of $2 - 3 ~M_{\odot}$. This mass range implies that
the deposited energy of the jet ranges between
$0.5 \times 10^{52}$ to $2 \times 10^{52}$ erg. 
This also constrains the deposition time for this transient about $O(1)$ s. The final black hole mass will be $\sim 12 - 13 ~M_{\odot}$, depending on the 
exact details of the jet. 

\subsection{Ejecta energy against deposited energy}

\begin{figure}
\centering
\includegraphics*[width=8.5cm]{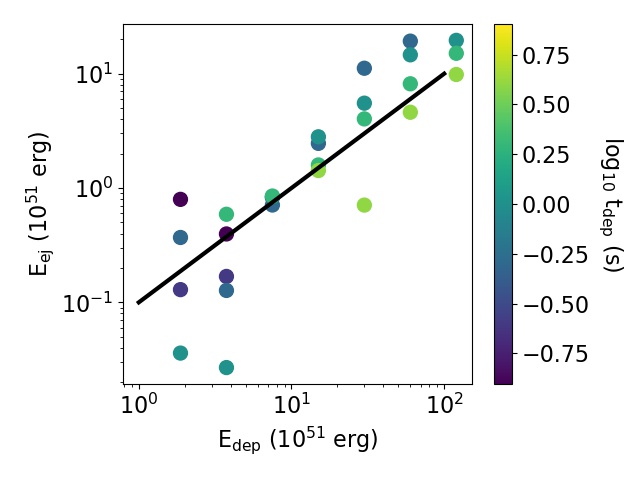}
\caption{The ejecta energy against deposited energy from the 
models presented in this work. The solid line corresponds to ratio 
$E_{\rm ejecta} = 0.1 E_{\rm dep}$.}
\label{fig:Eej_Edep}
\end{figure}

Due to the massive envelope and aspherical deposition of energy,
the typical efficiency of energy deposition is low. That means,
the energy contained in the ejecta, as the sum of
their internal and kinetic energy, can be much lower than the total
energy deposited by the jet. Here we examine how the two 
quantities are related. 

To obtain the ejecta energy, we compute the sum of ejecta kinetic energy
where the tracers have a positive sum of their kinetic energy and gravitational energy. In Figure \ref{fig:Eej_Edep} we plot the ejecta total 
energy against deposited energy by the jet. 
We also plot a line for $E_{\rm ejecta} = 0.1 ~E_{\rm dep}$.
The straight line, which corresponds to an efficiency of 10 \%,
shows the typical trend of the models. Dispersion occurs at both very low
and very high $E_{\rm jet}$. At the low energy limit, the statistical fluctuation becomes significant. While at the high energy limit, how the jet deposits energy affects the mass ejection. This demonstrates the non-linear interaction of how the shocked matter leads to mass ejection.

\subsection{$^{56}$Ni mass against Deposited Energy}

\begin{figure}
\centering
\includegraphics*[width=8.5cm]{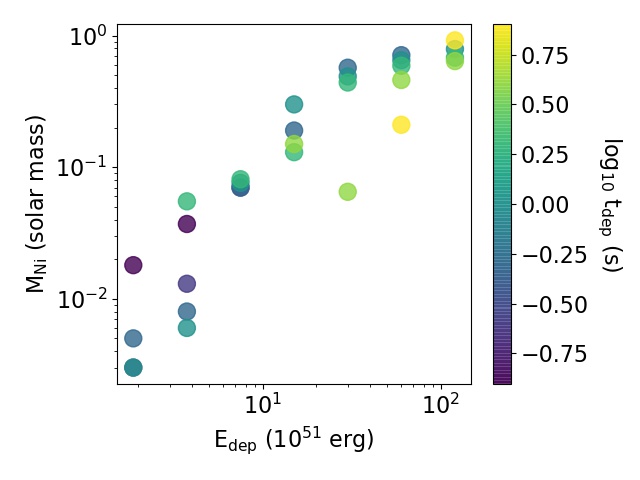}
\caption{The $^{56}$Ni mass against deposited energy for the 
models presented in this work.}
\label{fig:Ni56_dist}
\end{figure}

The hypernova SN 1998bw associated with GRB 980425 has shown
a possibility of triggering a jet from an aspherical
explosion.  Here we further examine the ejected $^{56}$Ni mass as a
function of the explosion energy. These quantities can be directly constrained by the light curves shape and the spectral lines. Here we explore the correlation of this pair of quantities. 

In Figure \ref{fig:Ni56_dist} we plot the final ejected $^{56}$Ni mass against the explosion energy. The explosion energy shows an almost monotonic relation with the $^{56}$Ni mass, which grows from $\sim 10^{-2} ~M_{\odot}$ to as high as 0.8 $M_{\odot}$ in the high energy limit. Different configurations
of the jet produce a dispersion of the $^{56}$Ni mass about 0.2 -- 0.4 $M_{\odot}$. 
Similar to the ejected mass, the ejecta $^{56}$Ni mass also levels off
at $4 \times 10^{51}$ erg. 

SN 1998bw has a derived $^{56}$Ni mass of $0.2 - 0.6 ~M_{\odot}$. Our models suggest that, in order to achieve the derived $^{56}$Ni mass, a minimum of explosion energy $> 10^{52}$ erg is necessary to reconcile with the lower limit 0.2 $M_{\odot}$. On the other hand, in the upper limit, $E_{{\rm dep}}$ from 2 to 8 $\times 10^{52}$ erg will be necessary.

\subsection{$^{56}$Ni mass against Ejecta velocity}

\begin{figure}
\centering
\includegraphics*[width=8.5cm]{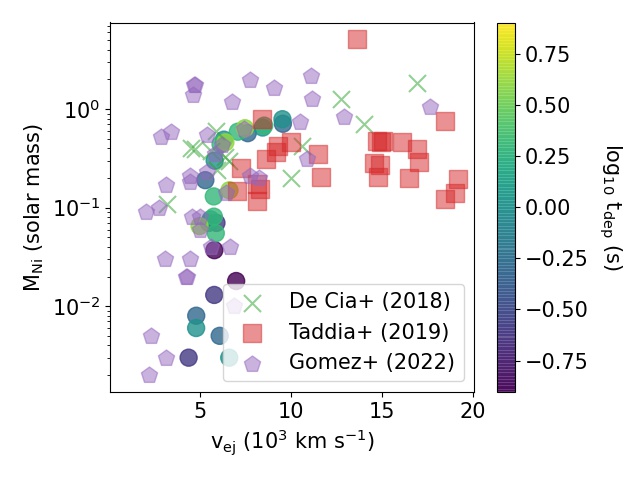}
\caption{The $^{56}$Ni mass against ejecta characteristic velocity
for the models presented in this work. Other data points are taken from superluminous SNe Ib/c \citep{DeCia2018}, SNe Ic-BL \citep{Taddia2019} and luminous SNe Ib/c \citep{Gomez2022}.}
\label{fig:Ni56_vej}
\end{figure}

Another important observable pair of a jet-driven supernova
is the $^{56}$Ni mass against typical ejecta velocity. This pair of variables
can be derived from the light curve peak luminosity $L_{{\rm peak}}$
against the Si II velocity inferred from the spectra. Here we 
estimate the typical ejecta velocity by first calculating the 
ejecta total energy $E_{\rm ej}$ and its mass $M_{\rm ej}$.
The typical ejecta velocity is defined as $v_{\rm ej} = \sqrt{2 E_{\rm ej}/M_{\rm ej}}.$

In Figure \ref{fig:Ni56_vej} we plot the $^{56}$Ni mass against $v_{\rm ej}$ for the models presented in this work. The distribution of data points is more scattered than the $E_{\rm ej}$--$E_{\rm dep}$ pair. At a low $v_{\rm ej}$, the two quantities have a linear relation but with a large dispersion. This trend extends from $4 - 7 \times 10^3$ km s$^{-1}$. At a high $v_{\rm ej}$, the $^{56}$Ni mass is almost independent of the ejecta velocity. 

The estimation does not fully capture the details of the ejecta velocity because the the outer layer of ejecta can move faster than the bulk of ejecta, after the ejecta develops a homologous expansion profile. The deposition of the radioactive decay energy of $^{56}$Ni by $\gamma$-ray may also create secondary changes to the asymptotic velocity profile \citep{Blinnikov2006}. Thus the velocity here represents the average velocity of the total ejecta.

To further compare the models, we compare out models with the observational results from superluminous SNe Ib/c \citep{DeCia2018}, SNe Ic-BL \citep{Taddia2019} and luminous SNe Ib/c \citep{Gomez2022}. The parameters are taken from their fitting results using analytical models. The observational data also shows a knee structure as in our models, with low ejecta mass spanning a wide range of $^{56}$Ni from $10^{-2} - 10^{0}~M_{\odot}$. Meanwhile, models with a high $v_{\rm ej}$ correspond to a smaller range but a higher $^{56}$Ni mass value. Some supernovae shows a higher $v_{\rm ej} \sim 2\times 10^4$ km s$^{-1}$. This is likely to be aspherical effects where the ejected $^{56}$Ni has a higher velocity than the bulk of ejecta.

\subsection{$^{56}$Ni mass against Ejecta Mass}

\begin{figure}
\centering
\includegraphics*[width=8.5cm]{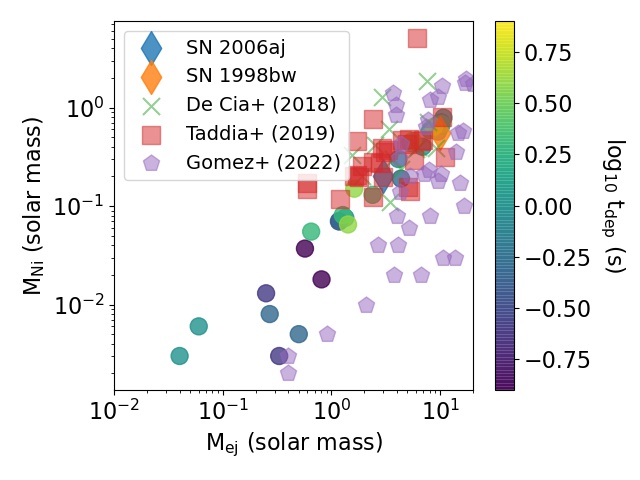}
\caption{The $^{56}$Ni mass in the ejecta against ejecta
mass from the models presented in this work. Observational
data from SN 1998bw and SN 2006aj are included as 
a comparison \citep{Nomoto2010}. Observational data are taken similarly from \cite{DeCia2018, Taddia2019, Gomez2022} as in Figure \ref{fig:Ni56_vej}.}
\label{fig:Mni_Mej}
\end{figure}

Another observable pair presented is the $^{56}$Ni mass, $M_{\rm Ni}$ against the ejecta
mass $M_{\rm ej}$. These two variables can be directly extracted from the shape and peak luminosity of the light curves. 

In Figure \ref{fig:Mni_Mej} we plot the $^{56}$Ni mass against the total ejecta mass for the models in this work. Our data shows a strong correlation between this pair of observables. The quantity pair exhibits a power-law scaling. A lower $M_{\rm ej}$ leads to a larger dispersion in $M_{\rm Ni}$ and the dispersion decreases in the high $M_{\rm ej}$ limit. 
We also use the observational data to demonstrate how our models can indicate the explosion history of observed hypernovae. 

SN 1998bw is the first evidence for GRB-hypernova correlation \citep{Galama1998}. The supernova is modeled as an explosion of a massive CO star \citep{Iwamoto1998,Woosley1999} with a fit of $^{56}$Ni $\sim$ 0.5 $M_{\odot}$ and $M_{\rm ej} \sim 10$ $M_{\odot}$. 
SN 2006aj is regarded as a hypernova explosion on the dim side. 
Fitting of the spectra and light curve gives an estimate
$M_{\rm ejecta}  = 3$ $M_{\odot}$ and $^{56}$Ni $\sim$ 0.2 $M_{\odot}$.

Both models favour models with $t_{\rm jet} > t_{\rm jet,0}$ to fit this observable pair. These data points show that the observed hypernovae are diversified in the parameter space of jet energetics. Our models agree with the trend derived from observed hypernovae.

We further compare the statistical trend of Type Ib/c(-BL) supernovae which are in the superluminous and luminous branch. Our models show the slope of this variable pair being consistent with the slope and the mass range presented in their derived models, especially the SNe Ic/BL models in \cite{Taddia2019}. 

However, there are a few supernovae from \cite{Gomez2022} with $M_{\rm Ni}$ below $\sim 10^{-2}$ where our models are persistently higher than theirs. They correspond to a very low energy deposition which we expect the interpretation of $^{56}$Ni could encounter greater uncertainties due to late-time fallback and low statistics.

\subsection{Sc-Ti-V correlation}

\begin{figure}
\centering
\includegraphics*[width=8.5cm]{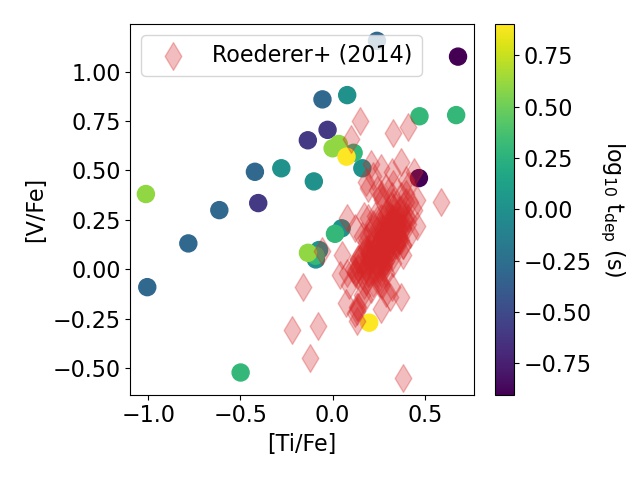}
\caption{[V/Fe] against [Ti/Fe] for the models presented in this work.
The data points are taken from the metal poor star survey reported in \cite{Roederer2014}. Only stars with a determined values of [Ti/Fe] and [V/Fe] are plotted.}
\label{fig:Ti_V_correlation}
\end{figure}

Here we study how the Sc-Ti-V correlation depends on 
our models. This correlation has been observed in metal-poor stars \citep{Sneden2016}. The metal-poor star HD 84937, together with others, have shown that the three element ratios are correlated. The amount of Ti increases with V and Sc. Here we focus on Ti-V relation.

To obtain the two quantities [Ti/Fe] and [V/Fe], we use the chemical composition of the ejected particles. Then, we compute the corresponding Ti and V masses. In Figure \ref{fig:Ti_V_correlation} we plot [V/Fe] against [Ti/Fe] for the models presented in this work. In \cite{Sneden2016} this pair of elements shows a correlation of  45$^{\circ}$ based on the metal-poor stars derived in \cite{Roederer2014}. We overlay the stellar abundance data on our models. 

There is an intrinsic scattering in the model prediction due to the nonlinear dependence of the deposited energy on how the ejecta is heated and how much matter is ejected. Some sequences of
models, at such a time as $t_{\rm jet} = 0.5 t_{\rm jet,0}$, 
show a clear trend of how [Ti/Fe] scales with [V/Fe] which spans from $\sim -1.0$ to 0.5 in [Ti/Fe]. The trend becomes less obvious for other choices of $t_{\rm jet}$.
The trend of our models are comparable with the trend of observational data. Some of the models lie on the outskirt of the clusters. However, no model in our sequence directly intersects with the cluster. This can be attributed to the progenitor considered. The production of [Ti/Fe] in general peaks around a maximum temperature of $\sim 3.5 \times 10^9$ K. At a higher temperature, the production of Fe-group elements surpasses the production of Ti. As shown in Figure \ref{fig:summary_benchmark}, the jet excites matter to a maximum temperature above $5 \times 10^9$ K. Thus, the production of Fe-group elements dominates and suppresses the abundance ratios [Ti/Fe]. This suggests that to reconcile with the observational group, a jet of lower energy or a more compact (i.e. a lower mass) progenitor is necessary.

By comparing with Figure \ref{fig:Mni_Mej}, we observe that the high $\dot{E}_{\rm dep}$ models can simultaneously correspond to observed high [Ti/Fe] vs. [V/Fe] stars and hypernovae. However, the low $\dot{E}_{\rm dep}$ models are less relevant to the observed abundances in these stars. 

In our models, the Sc production is below the solar value, regardless of the use of $\nu-$p process in nuclear reactions. Our model shows some Sc production during the jet propagation but it is destroyed at later time. It suggests that the underproduction has a systematic dependence on the progenitor. Further extension with a realistic progenitor will be important to confirm this suggestion. 

\subsection{High Zn/Fe Stars}

Recent analysis of a few low metallicity stars shows an enhanced production of [Zn/Fe] from low-metallicity supernovae. These results suggest the aspherical explosion due to the local high entropy environment as in the 
electron-capture supernova and hypernova branch \citep{Nomoto2010}. Early works demonstrated the connection between
metal poor stars and aspherical explosions from the peculiar pattern of some carbon enhanced metal-poor stars, including HE1300+0157, HE1327-2326, HE0107-5240 and HE1424-0241 \cite [e.g.][]{Tominaga2009}. Some of these objects, in particular
HE1300+0157 and 
HE1327-2326, have [C/Fe] as high as 1 -- 4. 
To explain the low Fe abundance, the ejecta restricted
within certain angle is necessary to produce
these results.

\begin{figure}
\centering
\includegraphics*[width=8.5cm]{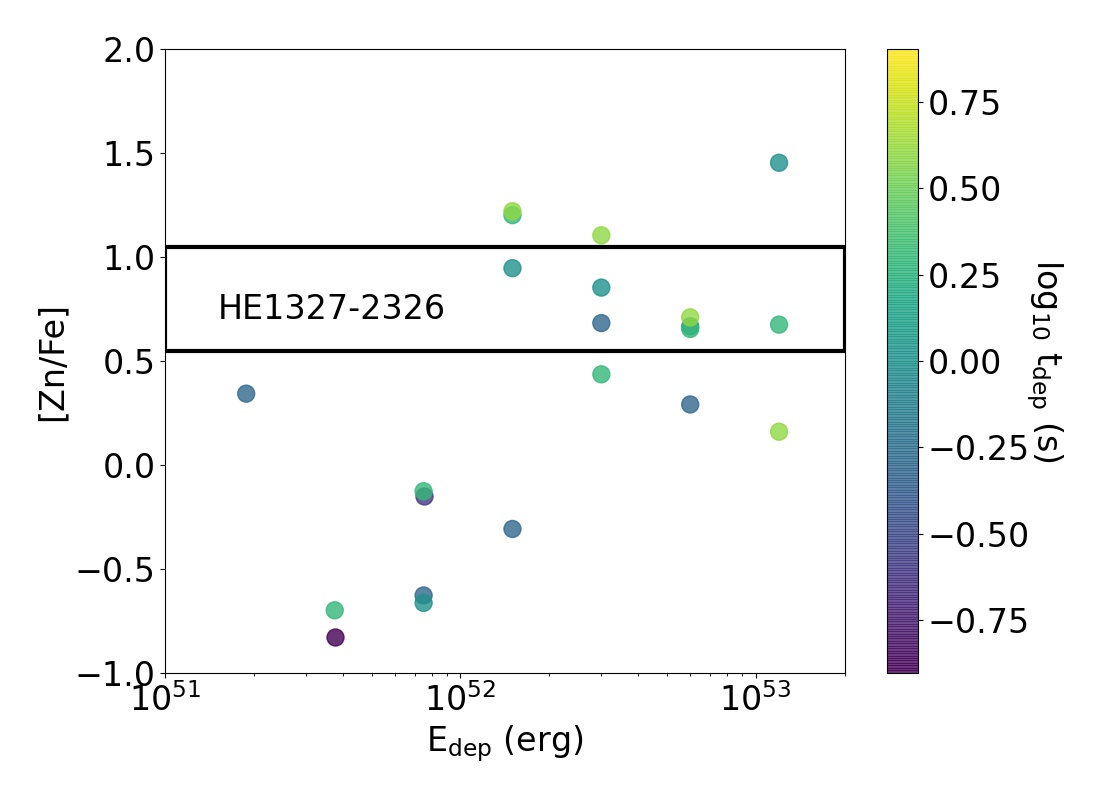}
\caption{The ejecta mass against explosion energy from the 
models presented in this work. The box corresponds to the derived abundance of Zn/Fe for HE1327-2326 \citep{Ezzeddine2019}.}
\label{fig:Zn_dist}
\end{figure}

In Figure \ref{fig:Zn_dist}, we compare the ratio [Zn/Fe] taken from our models as a function of the explosion energy. At the low energy limit, the ratio [Zn/Fe] is suppressed for two reasons. First, the jet deposits insufficient energy, where the heated matter in the core does not have
the necessary momentum to breakthrough the heavy envelope. 
Second, at this low energy, the ejecta does not go through the high-entropy phase. The temperature of the ejecta cannot reach NSE for $^{56}$Ni-production or further. 
On the other hand, in the high energy limit, the upper limit of [Zn/Fe] in our models can always reach the values suggested by observational data. Scatter of [Zn/Fe] exists for a given $E_{\rm dep}$. It is because the duration of energy deposition can also affect the ejected mass and the production of $^{56}$Fe, which alter the ratio directly.

We compare our models with the recent observations of HE1327-2326. This star shows a significant enhancement of [Zn/Fe] = $0.8 \pm 0.25$. Such a high ratio indicates the high entropy environment typical in jet-driven supernova is necessary. In order to realize the high [Zn/Fe] in HE1327-2326, we observe that  
from our models, $E_{\rm dep} = 10^{52} - 4 \times 10^{52}$ erg is necessary for producing the expected [Zn/Fe]. In Tables 
\ref{table:abundance_table} -- \ref{table:radioactive_table}
we present the chemical abundance tables for the stable and radioactive 
isotopes in the ejecta of our collapsar models where the [Zn/Fe] ratio satisfies the observed value. The results agree that the high-entropy environment for the $\alpha$-rich freezeout is essential for the high-Zn production.

\subsection{Caveats}
\label{sec:caveats}

In this work we have considered only the 40 $M_{\odot}$ zero-metallicity star as the progenitor. We do not consider models of other masses due to the expensive computation of these multi-dimensional models. For lower mass models, there could be substantial fall back accretion before the black hole forms. The infalling matter will be distinctive from our models. On the other hand, we expect less changes appear for higher mass models as the progenitor structure does not vary qualitatively. The Si- and C+O-cores are in general non-degenerate. To a good approximation, extending this approach to both higher and lower mass star models may demonstrate the diversity of jet-driven supernova.

In this work we also assumed the progenitor stellar model to be spherical symmetric and non-rotating. Indeed, the formation of gamma-ray bursts through a black hole accretion disk requires an initial angular momentum. Since we are not directly modeling how the infall of matter triggers the outburst of energy, the rotation component is less important. Furthermore, the rotation in typical stars is less than $\sim 10\%$ of the critical rotation. The dynamical effect of rotation on the ejecta is small. One possible effect by the neglect of rotation in the stellar model is the difference in the outer layer where rotational mixing is important during main-sequence evolution. However, as most Fe-group element synthesis is relevant in the Si- and inner C+O layer, rotation plays a less important role. Despite that, the study of how rotation is coupled to stellar evolution and the explosion will be important for future quantitative comparisons.

\subsection{Conclusion}

In this work we have examined the parameter dependence of the hydrodynamics and 
nucleosynthesis of jet-driven supernovae.
We use the 40 $M_{\odot}$ zero-metallicity star at the onset of 
Fe-core collapse as the progenitor with a mass cut at $\sim 1.4 ~M_{\odot}$.
We treat the energetics of the jet, including the energy deposition
rate, energy deposition time and the jet open-angle. We vary the total 
deposited energy ranging from $\sim 10^{50}$ to $\sim 10^{52}$ erg.

We observe the following features in the explosion morphology and 
nucleosynthesis pattern. 

\noindent (1) The ejecta mass is sensitive to the energy deposition rate and the energy deposition duration. A large energy deposition can trigger a mass
ejection along the jet-open angle and the outermost CO and He layers. The ejecta composition features the presence of Ni, Cu and Zn in general. 

\noindent (2) The ejected mass is also sensitive to the jet open angle. A wider jet open angle results in a more dispersed energy deposition. There is less matter along the jet open-angle and more matter in the outer layer. The ejecta in a wide angle jet contains more light elements including C, O, Ne and IMEs (e.g. Si, S and Ar). The ejecta contains much fewer IGEs, especially Ni, Cu and Zn. Observational data of metal poor stars in Ti, V, Cr for the light IGEs and Ni, Co, Zn for the heavy IGEs can constrain 
directly the explosion energetics.

\noindent (3) The corresponding explosion results in a moderately
bright event compared to Type Ic supernovae. Typical
collapsar explosions produce 0.1 -- 0.3 $M_{\odot}$ of $^{56}$Ni. 
In terms of the ejecta mass, a weaker explosion can eject matter as low as $\sim$ 0.01 $M_{\odot}$,
while a strong explosion can eject $\sim 10 ~M_{\odot}$. 
The effective energy deposition in the ejecta is $\sim$ 10\% of the actual deposited energy. The observed energy of the ejecta could therefore be a probe of the initial energy deposition by the jet, which constrains the jet formation environment.
 
\noindent (4) The ejected $^{56}$Ni mass is strongly sensitive to the deposited energy. It ranges from $\sim 10^{-3} ~M_{\odot}$ in the supernova regime and $\sim 1 ~M_{\odot}$ in the hypernova regime. The corresponding $^{56}$Ni mass against $v_{\rm ej}$ shows a knee pattern. The lower $v_{\rm ej}$ grows with $^{56}$Ni mass, while at high $v_{\rm ej}$ $^{56}$Ni mass is capped above. 

\noindent (5) We examine the relation between [V/Fe] and [Ti/Fe] of our models. We show that our jet-driven supernovae can reproduce the 45 degree of slope derived from the observed metal-poor stars \citep{Roederer2014,Sneden2016}. This connects with the
metal-poor star statistics that the observed abundance pattern originates from early jet-driven supernovae. However, an exact matching will require the extension to other progenitor models.

\noindent (6) We compare the model abundance patterns with the recently observed Zn-enriched metal-poor star HE 1327+2326. The high Zn ratio coincides with our jet-driven supernova models with an explosion energy $\sim 1\times 10^{52}$ erg. Future observations and statistics of this abundance ratio from metal-poor stars may provide important constraints on the jet energetics.

\section{Acknowledgement}

S.C.L. acknowledges support by funding HST-AR-15021.001-A and 80NSSC18K1017.
K.N. acknowledges support by World Premier International Research Center Initiative (WPI), and JSPS KAKENHI Grant Numbers JP17K05382, JP20K04024, JP21H04499 and JP23K03452.

We thank Frank Timmes for the open-source subroutines of
the Helmholtz equation and state and the 
torch nuclear reaction network. 
We thank Nozomu Tominaga for the background and details 
in how jet-driven supernova is modeled. 
We thank Sachiko Tsuruta for the interesting introduction
in the mechanism of launch a jet from a compact stellar object. 
We thank Rana Ezzeddine for the inspiring 
discussion on the Zn-rich metal-poor star.

\newpage

\begin{table*}
\begin{center}
\caption{The abundance tables for the jet-driven supernova models where the 
[Zn/Fe] satisfies the observed value of the extreme metal-poor star. All short-lived
isotopes are assumed to have decayed. }

\begin{tabular}{ c c c c c c c }

\hline
Isotope & S40-4000-0500-15 & S40-1000-1000-15 & S40-2000-1000-15 & S40-4000-1000-15 & S40-0500-4000-15 & S40-1000-4000-15\\ \hline
$^{12}$C & $3.88 \times 10^{-1}$ & $1.89 \times 10^{-1}$ & $3.27 \times 10^{-1}$ & $3.82 \times 10^{-1}$ & $2.17 \times 10^{-2}$ & $2.48 \times 10^{-1}$ \\
 $^{13}$C & $2.47 \times 10^{-9}$ & $7.0 \times 10^{-10}$ & $1.31 \times 10^{-9}$ & $2.56 \times 10^{-9}$ & $2.17 \times 10^{-10}$ & $9.36 \times 10^{-10}$ \\
 $^{14}$N & $4.67 \times 10^{-7}$ & $4.32 \times 10^{-7}$ & $4.72 \times 10^{-7}$ & $4.62 \times 10^{-7}$ & $3.82 \times 10^{-7}$ & $4.57 \times 10^{-7}$ \\
 $^{15}$N & $8.61 \times 10^{-7}$ & $7.14 \times 10^{-7}$ & $7.54 \times 10^{-7}$ & $8.59 \times 10^{-7}$ & $6.30 \times 10^{-7}$ & $7.53 \times 10^{-7}$ \\
 $^{16}$O & $45.93 \times 10^{-1}$ & $14.81 \times 10^{-1}$ & $30.17 \times 10^{-1}$ & $49.28 \times 10^{-1}$ & $1.46 \times 10^{-1}$ & $20.64 \times 10^{-1}$ \\
 $^{17}$O & $1.9 \times 10^{-9}$ & $5.89 \times 10^{-10}$ & $6.76 \times 10^{-10}$ & $1.89 \times 10^{-9}$ & $3.63 \times 10^{-10}$ & $6.27 \times 10^{-10}$ \\
 $^{18}$O & $2.2 \times 10^{-7}$ & $2.3 \times 10^{-7}$ & $2.4 \times 10^{-7}$ & $1.94 \times 10^{-7}$ & $1.62 \times 10^{-7}$ & $2.4 \times 10^{-7}$ \\
 $^{19}$F & $1.29 \times 10^{-9}$ & $2.54 \times 10^{-10}$ & $2.63 \times 10^{-10}$ & $1.10 \times 10^{-9}$ & $7.87 \times 10^{-11}$ & $1.95 \times 10^{-10}$ \\
 $^{20}$Ne & $2.40 \times 10^{-1}$ & $1.0 \times 10^{-1}$ & $1.92 \times 10^{-1}$ & $2.39 \times 10^{-1}$ & $5.83 \times 10^{-3}$ & $1.41 \times 10^{-1}$ \\
 $^{21}$Ne & $8.31 \times 10^{-6}$ & $2.95 \times 10^{-6}$ & $5.41 \times 10^{-6}$ & $8.43 \times 10^{-6}$ & $4.69 \times 10^{-7}$ & $3.93 \times 10^{-6}$ \\
 $^{22}$Ne & $5.37 \times 10^{-6}$ & $3.14 \times 10^{-6}$ & $4.9 \times 10^{-6}$ & $5.38 \times 10^{-6}$ & $9.81 \times 10^{-7}$ & $3.56 \times 10^{-6}$ \\
 $^{23}$Na & $8.69 \times 10^{-5}$ & $2.71 \times 10^{-5}$ & $4.71 \times 10^{-5}$ & $8.78 \times 10^{-5}$ & $5.5 \times 10^{-6}$ & $3.18 \times 10^{-5}$ \\
 $^{24}$Mg & $2.0 \times 10^{-1}$ & $6.22 \times 10^{-2}$ & $1.17 \times 10^{-1}$ & $2.20 \times 10^{-1}$ & $6.19 \times 10^{-3}$ & $8.25 \times 10^{-2}$ \\
 $^{25}$Mg & $1.99 \times 10^{-4}$ & $6.19 \times 10^{-5}$ & $1.13 \times 10^{-4}$ & $2.12 \times 10^{-4}$ & $8.11 \times 10^{-6}$ & $7.91 \times 10^{-5}$ \\
 $^{26}$Mg & $7.90 \times 10^{-5}$ & $2.11 \times 10^{-5}$ & $4.28 \times 10^{-5}$ & $8.67 \times 10^{-5}$ & $2.87 \times 10^{-6}$ & $3.2 \times 10^{-5}$ \\
 $^{26}$Al & $2.48 \times 10^{-28}$ & $1.65 \times 10^{-27}$ & $1.79 \times 10^{-28}$ & $4.48 \times 10^{-27}$ & $4.64 \times 10^{-28}$ & $1.89 \times 10^{-27}$ \\
 $^{27}$Al & $5.60 \times 10^{-3}$ & $1.73 \times 10^{-3}$ & $3.18 \times 10^{-3}$ & $6.13 \times 10^{-3}$ & $1.80 \times 10^{-4}$ & $2.22 \times 10^{-3}$ \\
 $^{28}$Si & $2.99 \times 10^{-1}$ & $5.93 \times 10^{-2}$ & $1.60 \times 10^{-1}$ & $3.80 \times 10^{-1}$ & $2.2 \times 10^{-2}$ & $1.54 \times 10^{-1}$ \\
 $^{29}$Si & $1.0 \times 10^{-3}$ & $2.50 \times 10^{-4}$ & $5.37 \times 10^{-4}$ & $1.24 \times 10^{-3}$ & $3.85 \times 10^{-5}$ & $3.80 \times 10^{-4}$ \\
 $^{30}$Si & $2.35 \times 10^{-3}$ & $8.8 \times 10^{-4}$ & $1.67 \times 10^{-3}$ & $2.83 \times 10^{-3}$ & $1.0 \times 10^{-4}$ & $1.44 \times 10^{-3}$ \\
 $^{31}$P & $6.94 \times 10^{-4}$ & $2.73 \times 10^{-4}$ & $4.90 \times 10^{-4}$ & $8.54 \times 10^{-4}$ & $3.74 \times 10^{-5}$ & $5.48 \times 10^{-4}$ \\
 $^{32}$S & $1.43 \times 10^{-1}$ & $2.59 \times 10^{-2}$ & $8.0 \times 10^{-2}$ & $1.86 \times 10^{-1}$ & $1.16 \times 10^{-2}$ & $8.72 \times 10^{-2}$ \\
 $^{33}$S & $1.78 \times 10^{-4}$ & $4.6 \times 10^{-5}$ & $9.2 \times 10^{-5}$ & $2.31 \times 10^{-4}$ & $7.22 \times 10^{-6}$ & $6.67 \times 10^{-5}$ \\
 $^{34}$S & $9.69 \times 10^{-4}$ & $3.63 \times 10^{-4}$ & $6.67 \times 10^{-4}$ & $1.13 \times 10^{-3}$ & $4.20 \times 10^{-5}$ & $6.7 \times 10^{-4}$ \\
 $^{36}$S & $1.50 \times 10^{-8}$ & $4.76 \times 10^{-9}$ & $8.70 \times 10^{-9}$ & $1.88 \times 10^{-8}$ & $3.26 \times 10^{-10}$ & $6.43 \times 10^{-9}$ \\
 $^{35}$Cl & $2.31 \times 10^{-4}$ & $1.40 \times 10^{-4}$ & $1.98 \times 10^{-4}$ & $2.82 \times 10^{-4}$ & $9.64 \times 10^{-6}$ & $3.13 \times 10^{-4}$ \\
 $^{37}$Cl & $1.26 \times 10^{-5}$ & $3.43 \times 10^{-6}$ & $5.58 \times 10^{-6}$ & $1.72 \times 10^{-5}$ & $8.60 \times 10^{-7}$ & $4.99 \times 10^{-6}$ \\
 $^{36}$Ar & $2.96 \times 10^{-2}$ & $5.38 \times 10^{-3}$ & $1.68 \times 10^{-2}$ & $3.88 \times 10^{-2}$ & $2.35 \times 10^{-3}$ & $1.91 \times 10^{-2}$ \\
 $^{38}$Ar & $2.30 \times 10^{-4}$ & $8.96 \times 10^{-5}$ & $1.15 \times 10^{-4}$ & $2.79 \times 10^{-4}$ & $1.62 \times 10^{-5}$ & $1.13 \times 10^{-4}$ \\
 $^{40}$Ar & $1.47 \times 10^{-10}$ & $7.37 \times 10^{-10}$ & $2.37 \times 10^{-10}$ & $1.40 \times 10^{-10}$ & $1.27 \times 10^{-10}$ & $2.51 \times 10^{-10}$ \\
 $^{39}$K & $5.23 \times 10^{-5}$ & $4.9 \times 10^{-5}$ & $3.62 \times 10^{-5}$ & $6.46 \times 10^{-5}$ & $3.17 \times 10^{-6}$ & $5.38 \times 10^{-5}$ \\
 $^{40}$K & $2.72 \times 10^{-9}$ & $1.34 \times 10^{-8}$ & $2.5 \times 10^{-9}$ & $3.23 \times 10^{-9}$ & $6.69 \times 10^{-10}$ & $6.11 \times 10^{-9}$ \\
 $^{41}$K & $3.98 \times 10^{-6}$ & $1.5 \times 10^{-6}$ & $1.62 \times 10^{-6}$ & $4.67 \times 10^{-6}$ & $3.3 \times 10^{-7}$ & $1.43 \times 10^{-6}$ \\
 $^{40}$Ca & $3.2 \times 10^{-2}$ & $6.14 \times 10^{-3}$ & $1.79 \times 10^{-2}$ & $3.94 \times 10^{-2}$ & $2.28 \times 10^{-3}$ & $2.1 \times 10^{-2}$ \\
 $^{42}$Ca & $1.5 \times 10^{-5}$ & $6.57 \times 10^{-6}$ & $8.5 \times 10^{-6}$ & $1.23 \times 10^{-5}$ & $1.58 \times 10^{-6}$ & $5.78 \times 10^{-6}$ \\
 $^{43}$Ca & $1.4 \times 10^{-5}$ & $5.32 \times 10^{-6}$ & $1.5 \times 10^{-5}$ & $1.4 \times 10^{-5}$ & $2.34 \times 10^{-6}$ & $9.20 \times 10^{-6}$ \\
 $^{44}$Ca & $1.13 \times 10^{-4}$ & $5.86 \times 10^{-5}$ & $1.18 \times 10^{-4}$ & $1.6 \times 10^{-4}$ & $1.2 \times 10^{-5}$ & $8.84 \times 10^{-5}$ \\
 $^{46}$Ca & $1.3 \times 10^{-13}$ & $3.24 \times 10^{-9}$ & $1.55 \times 10^{-11}$ & $1.39 \times 10^{-9}$ & $1.15 \times 10^{-14}$ & $1.21 \times 10^{-11}$ \\
 $^{48}$Ca & $2.7 \times 10^{-19}$ & $2.54 \times 10^{-12}$ & $2.56 \times 10^{-10}$ & $2.91 \times 10^{-11}$ & $3.64 \times 10^{-20}$ & $8.44 \times 10^{-15}$ \\
 $^{45}$Sc & $1.29 \times 10^{-6}$ & $1.16 \times 10^{-6}$ & $1.19 \times 10^{-6}$ & $1.60 \times 10^{-6}$ & $2.54 \times 10^{-7}$ & $7.46 \times 10^{-7}$ \\
 $^{46}$Ti & $9.37 \times 10^{-5}$ & $8.19 \times 10^{-5}$ & $1.2 \times 10^{-4}$ & $9.56 \times 10^{-5}$ & $4.71 \times 10^{-5}$ & $1.48 \times 10^{-4}$ \\
 $^{47}$Ti & $6.45 \times 10^{-5}$ & $4.9 \times 10^{-5}$ & $6.92 \times 10^{-5}$ & $6.1 \times 10^{-5}$ & $2.73 \times 10^{-5}$ & $7.10 \times 10^{-5}$ \\
 $^{48}$Ti & $1.22 \times 10^{-3}$ & $5.83 \times 10^{-4}$ & $1.19 \times 10^{-3}$ & $1.20 \times 10^{-3}$ & $8.71 \times 10^{-5}$ & $8.88 \times 10^{-4}$ \\
 $^{49}$Ti & $2.44 \times 10^{-5}$ & $1.73 \times 10^{-5}$ & $2.3 \times 10^{-5}$ & $2.66 \times 10^{-5}$ & $3.1 \times 10^{-6}$ & $1.83 \times 10^{-5}$ \\
 $^{50}$Ti & $3.48 \times 10^{-11}$ & $1.14 \times 10^{-7}$ & $3.29 \times 10^{-10}$ & $1.68 \times 10^{-11}$ & $1.62 \times 10^{-12}$ & $6.66 \times 10^{-9}$ \\
 $^{50}$V & $8.22 \times 10^{-9}$ & $4.76 \times 10^{-7}$ & $1.11 \times 10^{-8}$ & $3.59 \times 10^{-9}$ & $3.40 \times 10^{-9}$ & $2.38 \times 10^{-7}$ \\
 $^{51}$V & $2.37 \times 10^{-4}$ & $2.56 \times 10^{-4}$ & $2.42 \times 10^{-4}$ & $2.50 \times 10^{-4}$ & $8.23 \times 10^{-5}$ & $3.39 \times 10^{-4}$ \\
 
 \end{tabular}
\label{table:abundance_table}
\end{center}
\end{table*}

\begin{table*}
\begin{center}
\caption{Continued from Table \ref{table:abundance_table}. }

\begin{tabular}{c c c c c c c}
 
 \hline
Isotope & S40-4000-0500-15 & S40-1000-1000-15 & S40-2000-1000-15 & S40-4000-1000-15 & S40-0500-4000-15 & S40-1000-4000-15 \\ \hline
$^{50}$Cr & $9.96 \times 10^{-5}$ & $1.68 \times 10^{-4}$ & $9.94 \times 10^{-5}$ & $9.45 \times 10^{-5}$ & $1.61 \times 10^{-5}$ & $1.69 \times 10^{-4}$ \\
 $^{52}$Cr & $8.71 \times 10^{-3}$ & $3.83 \times 10^{-3}$ & $7.44 \times 10^{-3}$ & $9.97 \times 10^{-3}$ & $6.78 \times 10^{-4}$ & $7.52 \times 10^{-3}$ \\
 $^{53}$Cr & $2.45 \times 10^{-4}$ & $1.59 \times 10^{-4}$ & $1.7 \times 10^{-4}$ & $2.32 \times 10^{-4}$ & $1.13 \times 10^{-5}$ & $1.94 \times 10^{-4}$ \\
 $^{54}$Cr & $5.64 \times 10^{-7}$ & $1.26 \times 10^{-5}$ & $1.38 \times 10^{-7}$ & $7.15 \times 10^{-8}$ & $5.50 \times 10^{-8}$ & $6.76 \times 10^{-6}$ \\
 $^{55}$Mn & $1.38 \times 10^{-3}$ & $4.15 \times 10^{-4}$ & $3.87 \times 10^{-4}$ & $6.28 \times 10^{-4}$ & $1.83 \times 10^{-4}$ & $7.76 \times 10^{-4}$ \\
 $^{54}$Fe & $2.42 \times 10^{-3}$ & $9.78 \times 10^{-4}$ & $6.35 \times 10^{-4}$ & $1.65 \times 10^{-3}$ & $1.65 \times 10^{-4}$ & $1.73 \times 10^{-3}$ \\
 $^{56}$Fe & $5.69 \times 10^{-1}$ & $2.97 \times 10^{-1}$ & $4.90 \times 10^{-1}$ & $6.48 \times 10^{-1}$ & $6.50 \times 10^{-2}$ & $4.63 \times 10^{-1}$ \\
 $^{57}$Fe & $1.14 \times 10^{-2}$ & $5.42 \times 10^{-3}$ & $8.35 \times 10^{-3}$ & $1.17 \times 10^{-2}$ & $1.74 \times 10^{-3}$ & $7.25 \times 10^{-3}$ \\
 $^{58}$Fe & $7.99 \times 10^{-7}$ & $1.12 \times 10^{-5}$ & $2.27 \times 10^{-7}$ & $1.45 \times 10^{-7}$ & $1.0 \times 10^{-6}$ & $3.93 \times 10^{-6}$ \\
 $^{60}$Fe & $1.43 \times 10^{-15}$ & $3.24 \times 10^{-7}$ & $3.26 \times 10^{-11}$ & $3.18 \times 10^{-10}$ & $1.29 \times 10^{-13}$ & $1.27 \times 10^{-11}$ \\
 $^{59}$Co & $9.32 \times 10^{-4}$ & $5.35 \times 10^{-4}$ & $9.45 \times 10^{-4}$ & $9.48 \times 10^{-4}$ & $1.95 \times 10^{-4}$ & $6.26 \times 10^{-4}$ \\
 $^{58}$Ni & $2.20 \times 10^{-2}$ & $7.80 \times 10^{-3}$ & $1.90 \times 10^{-2}$ & $2.21 \times 10^{-2}$ & $6.76 \times 10^{-3}$ & $1.5 \times 10^{-2}$ \\
 $^{60}$Ni & $1.48 \times 10^{-2}$ & $1.3 \times 10^{-2}$ & $1.61 \times 10^{-2}$ & $1.62 \times 10^{-2}$ & $3.90 \times 10^{-3}$ & $1.54 \times 10^{-2}$ \\
 $^{61}$Ni & $4.73 \times 10^{-4}$ & $4.12 \times 10^{-4}$ & $5.21 \times 10^{-4}$ & $5.13 \times 10^{-4}$ & $1.15 \times 10^{-4}$ & $3.70 \times 10^{-4}$ \\
 $^{62}$Ni & $3.97 \times 10^{-3}$ & $2.16 \times 10^{-3}$ & $3.34 \times 10^{-3}$ & $3.25 \times 10^{-3}$ & $5.96 \times 10^{-4}$ & $1.73 \times 10^{-3}$ \\
 $^{64}$Ni & $2.58 \times 10^{-7}$ & $2.99 \times 10^{-5}$ & $2.58 \times 10^{-7}$ & $1.93 \times 10^{-5}$ & $2.48 \times 10^{-7}$ & $4.48 \times 10^{-7}$ \\
 $^{63}$Cu & $1.44 \times 10^{-4}$ & $1.62 \times 10^{-4}$ & $2.16 \times 10^{-4}$ & $1.31 \times 10^{-4}$ & $6.66 \times 10^{-5}$ & $1.17 \times 10^{-4}$ \\
 $^{65}$Cu & $4.33 \times 10^{-5}$ & $6.62 \times 10^{-5}$ & $6.2 \times 10^{-5}$ & $4.40 \times 10^{-5}$ & $1.97 \times 10^{-5}$ & $4.65 \times 10^{-5}$ \\
 $^{64}$Zn & $3.83 \times 10^{-3}$ & $3.99 \times 10^{-3}$ & $5.9 \times 10^{-3}$ & $4.30 \times 10^{-3}$ & $9.42 \times 10^{-4}$ & $3.38 \times 10^{-3}$ \\
 $^{66}$Zn & $8.57 \times 10^{-4}$ & $3.7 \times 10^{-4}$ & $5.26 \times 10^{-4}$ & $7.73 \times 10^{-4}$ & $4.63 \times 10^{-4}$ & $6.52 \times 10^{-4}$ \\
 $^{67}$Zn & $1.83 \times 10^{-5}$ & $4.81 \times 10^{-5}$ & $2.76 \times 10^{-5}$ & $4.85 \times 10^{-5}$ & $9.54 \times 10^{-6}$ & $1.79 \times 10^{-5}$ \\
 $^{68}$Zn & $2.68 \times 10^{-4}$ & $4.46 \times 10^{-4}$ & $6.85 \times 10^{-4}$ & $2.94 \times 10^{-4}$ & $8.10 \times 10^{-5}$ & $2.24 \times 10^{-4}$ \\
 $^{70}$Zn & $7.66 \times 10^{-11}$ & $9.1 \times 10^{-6}$ & $4.86 \times 10^{-9}$ & $5.14 \times 10^{-5}$ & $4.37 \times 10^{-12}$ & $6.45 \times 10^{-9}$ \\ 
 \hline

\end{tabular}
\label{table:abundance_table2}
\end{center}
\end{table*}

\begin{table*}
\begin{center}

\caption{The radioactive isotopes for the jet-driven supernova models where the 
[Zn/Fe] satisfies the observed value of the extreme metal-poor star. }

\begin{tabular}{c c c c c c c }

 \hline
Isotope & S40-4000-0500-15 & S40-1000-1000-15 & S40-2000-1000-15 & S40-4000-1000-15 & S40-0500-4000-15 & S40-1000-4000-15 \\ \hline

 $^{22}$Na & $1.53 \times 10^{-6}$ & $4.16 \times 10^{-7}$ & $7.64 \times 10^{-7}$ & $1.59 \times 10^{-6}$ & $1.0 \times 10^{-7}$ & $5.36 \times 10^{-7}$ \\
 $^{26}$Al & $1.34 \times 10^{-5}$ & $2.8 \times 10^{-6}$ & $6.51 \times 10^{-6}$ & $1.74 \times 10^{-5}$ & $4.34 \times 10^{-8}$ & $5.48 \times 10^{-6}$ \\
 $^{39}$Ar & $3.5 \times 10^{-10}$ & $3.50 \times 10^{-9}$ & $2.73 \times 10^{-10}$ & $3.49 \times 10^{-10}$ & $1.30 \times 10^{-10}$ & $1.42 \times 10^{-9}$ \\
 $^{40}$K & $2.74 \times 10^{-9}$ & $1.35 \times 10^{-8}$ & $2.6 \times 10^{-9}$ & $3.25 \times 10^{-9}$ & $6.73 \times 10^{-10}$ & $6.14 \times 10^{-9}$ \\
 $^{41}$Ca & $3.44 \times 10^{-6}$ & $1.1 \times 10^{-6}$ & $1.50 \times 10^{-6}$ & $4.67 \times 10^{-6}$ & $2.87 \times 10^{-7}$ & $1.36 \times 10^{-6}$ \\
 $^{44}$Ti & $1.2 \times 10^{-4}$ & $5.48 \times 10^{-5}$ & $1.6 \times 10^{-4}$ & $9.96 \times 10^{-5}$ & $9.95 \times 10^{-6}$ & $8.61 \times 10^{-5}$ \\
 $^{48}$V & $1.81 \times 10^{-7}$ & $1.4 \times 10^{-6}$ & $3.81 \times 10^{-7}$ & $1.58 \times 10^{-7}$ & $2.41 \times 10^{-7}$ & $5.75 \times 10^{-7}$ \\
 $^{49}$V & $6.85 \times 10^{-7}$ & $6.2 \times 10^{-6}$ & $1.67 \times 10^{-6}$ & $5.11 \times 10^{-7}$ & $9.89 \times 10^{-7}$ & $2.76 \times 10^{-6}$ \\
 $^{53}$Mn & $2.13 \times 10^{-5}$ & $1.35 \times 10^{-4}$ & $4.36 \times 10^{-6}$ & $6.85 \times 10^{-6}$ & $1.43 \times 10^{-6}$ & $7.98 \times 10^{-5}$ \\
 $^{60}$Fe & $2.31 \times 10^{-14}$ & $4.66 \times 10^{-6}$ & $5.40 \times 10^{-10}$ & $4.58 \times 10^{-9}$ & $1.86 \times 10^{-12}$ & $1.83 \times 10^{-10}$ \\
 $^{56}$Co & $6.89 \times 10^{-5}$ & $2.32 \times 10^{-5}$ & $4.12 \times 10^{-6}$ & $1.52 \times 10^{-5}$ & $4.90 \times 10^{-6}$ & $4.53 \times 10^{-5}$ \\
 $^{57}$Co & $5.30 \times 10^{-5}$ & $7.13 \times 10^{-5}$ & $5.56 \times 10^{-6}$ & $1.39 \times 10^{-5}$ & $1.18 \times 10^{-5}$ & $6.97 \times 10^{-5}$ \\
 $^{60}$Co & $4.76 \times 10^{-10}$ & $4.14 \times 10^{-7}$ & $1.69 \times 10^{-9}$ & $2.87 \times 10^{-10}$ & $1.28 \times 10^{-8}$ & $1.90 \times 10^{-8}$ \\
 $^{56}$Ni & $5.69 \times 10^{-1}$ & $2.97 \times 10^{-1}$ & $4.90 \times 10^{-1}$ & $6.48 \times 10^{-1}$ & $6.50 \times 10^{-2}$ & $4.63 \times 10^{-1}$ \\
 $^{57}$Ni & $1.14 \times 10^{-2}$ & $5.34 \times 10^{-3}$ & $8.33 \times 10^{-3}$ & $1.17 \times 10^{-2}$ & $1.73 \times 10^{-3}$ & $7.18 \times 10^{-3}$ \\
 $^{59}$Ni & $2.87 \times 10^{-4}$ & $2.62 \times 10^{-4}$ & $4.78 \times 10^{-4}$ & $4.7 \times 10^{-4}$ & $1.27 \times 10^{-4}$ & $2.24 \times 10^{-4}$ \\
 $^{63}$Ni & $1.0 \times 10^{-7}$ & $2.95 \times 10^{-6}$ & $4.74 \times 10^{-8}$ & $9.22 \times 10^{-8}$ & $1.59 \times 10^{-7}$ & $1.39 \times 10^{-7}$ \\ \hline

\end{tabular}

\label{table:radioactive_table}
\end{center}
\end{table*}

\newpage

\appendix
\section{Code Test of the Special Relativistic Hydrodynamics Extension}

\begin{figure*}
\centering
\includegraphics*[width=5.7cm]{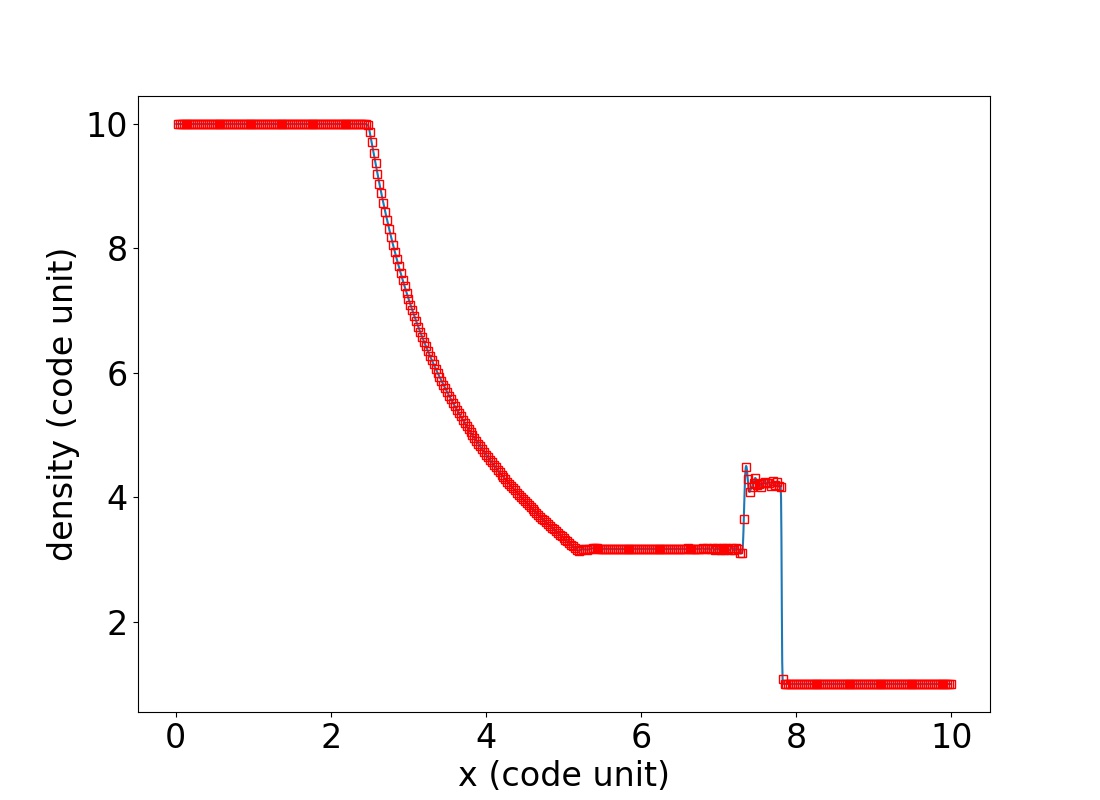}
\includegraphics*[width=5.7cm]{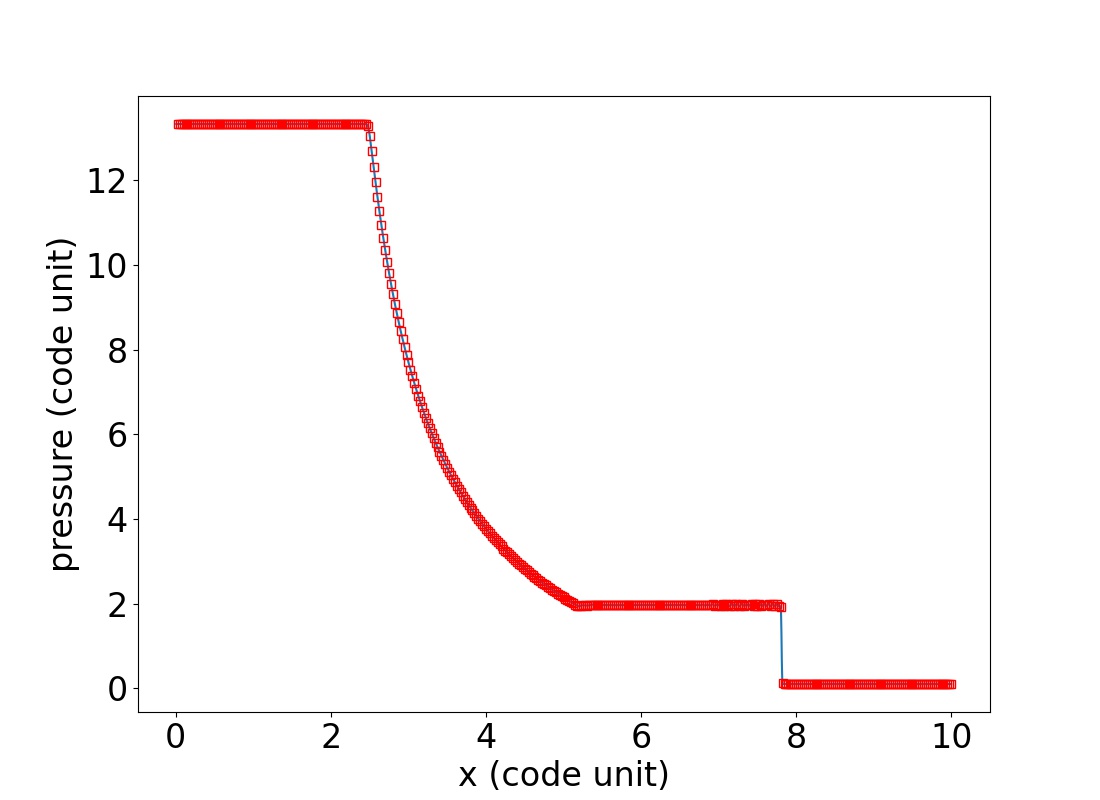}
\includegraphics*[width=5.7cm]{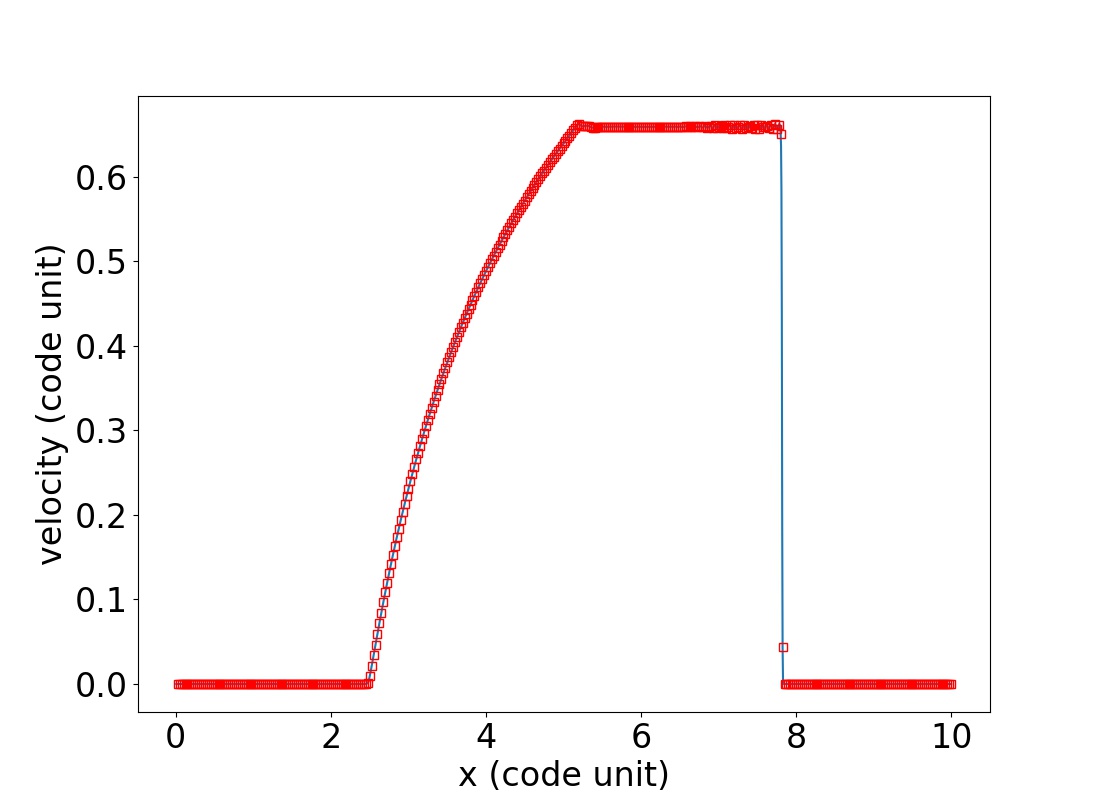}
\caption{(from left to right) The density, pressure and velocity
profiles of the Riemann problem test case taken from the Test 1 in \cite{Rosswog2009}. The red circles are the numerical data and the solid line is the analytic result.}
\label{fig:Rosswog1_test}
\end{figure*}

\begin{figure*}
\centering
\includegraphics*[width=5.7cm]{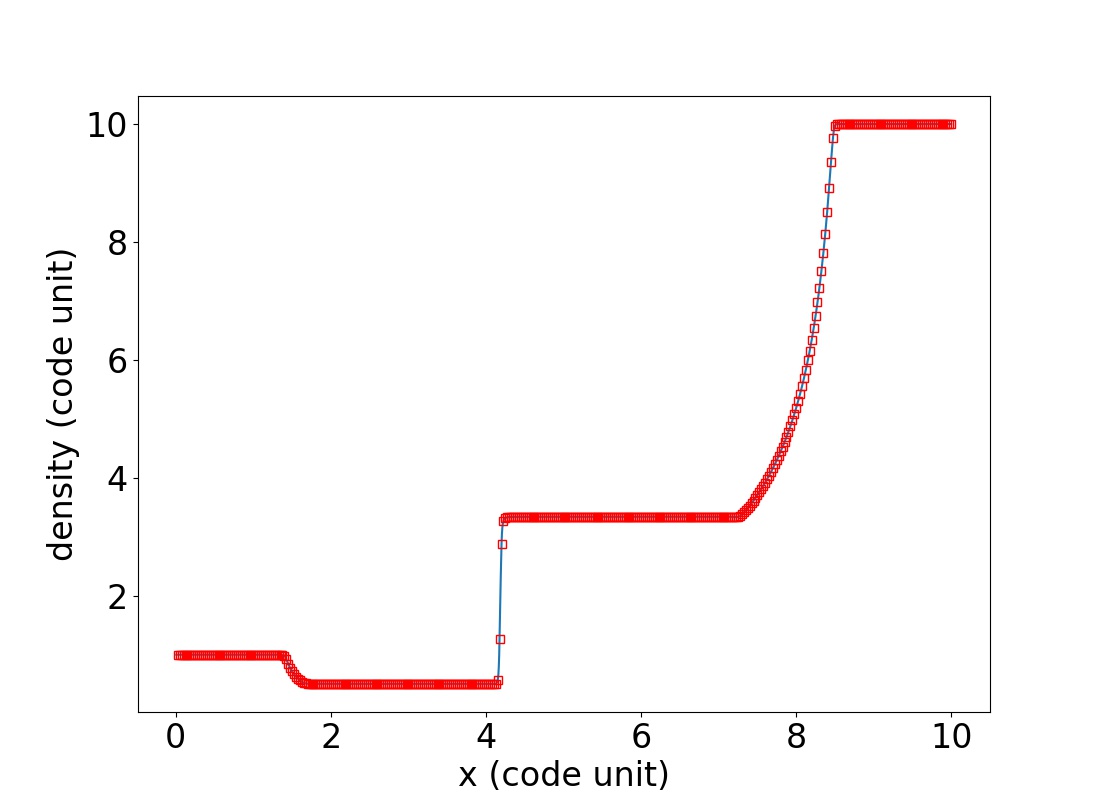}
\includegraphics*[width=5.7cm]{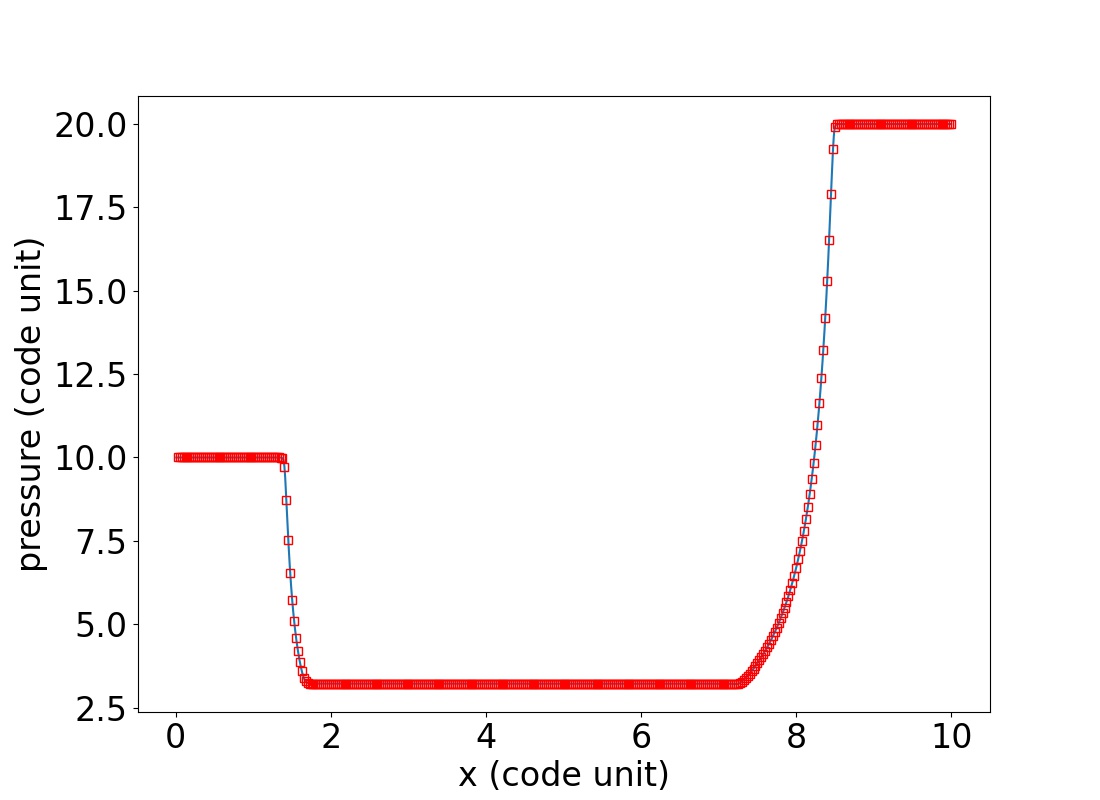}
\includegraphics*[width=5.7cm]{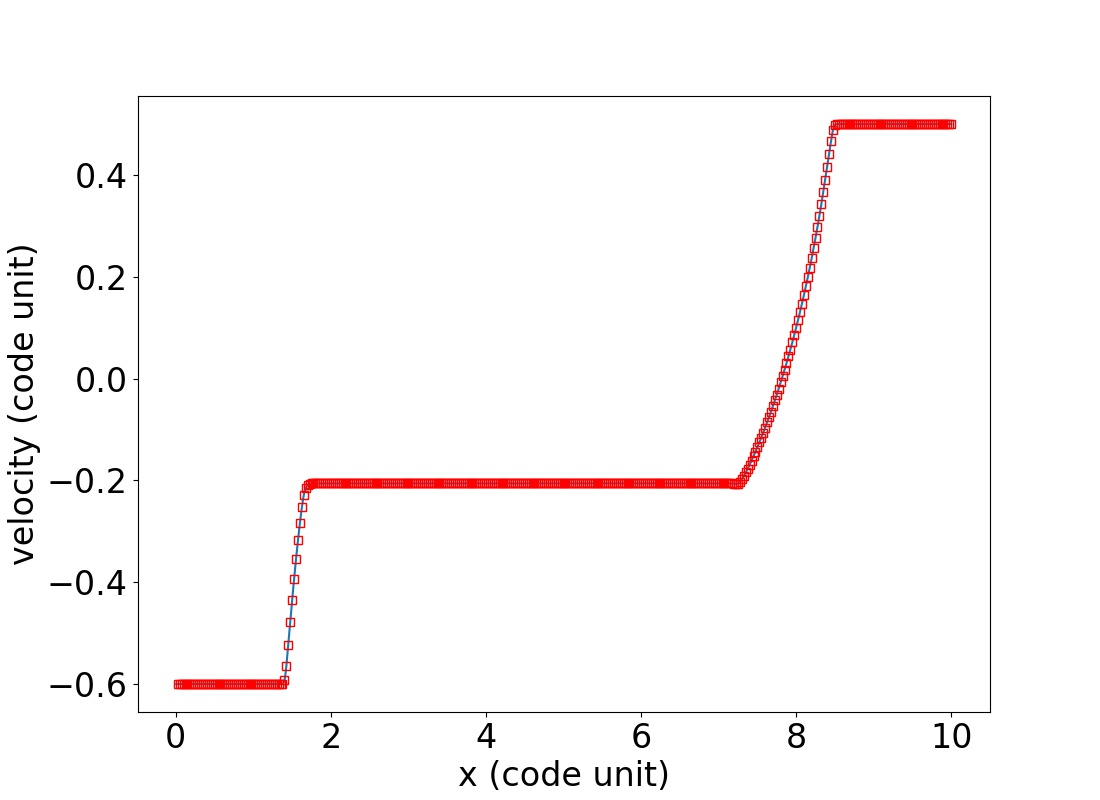}
\caption{Similar to Figure \ref{fig:Rosswog1_test} but from
the Riemann problem test case taken from the Test 2 in \cite{Marti1994}.}
\label{fig:Marti2_test}
\end{figure*}

\begin{figure*}
\centering
\includegraphics*[width=5.7cm]{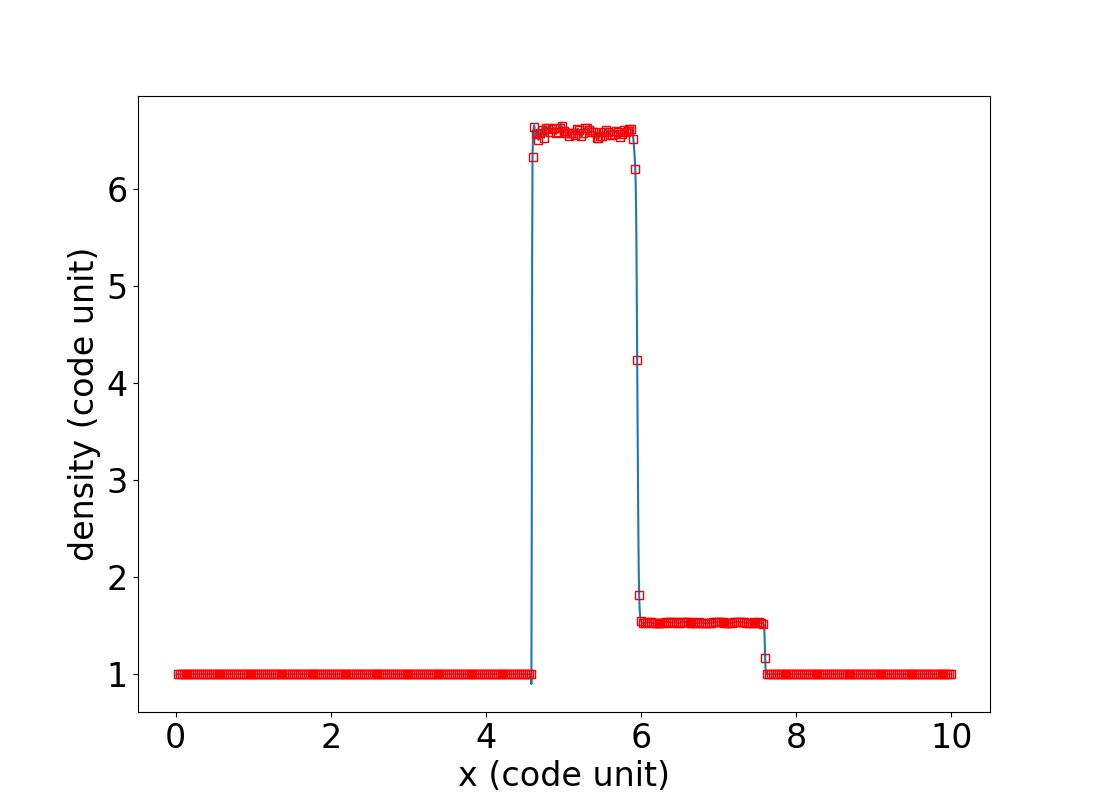}
\includegraphics*[width=5.7cm]{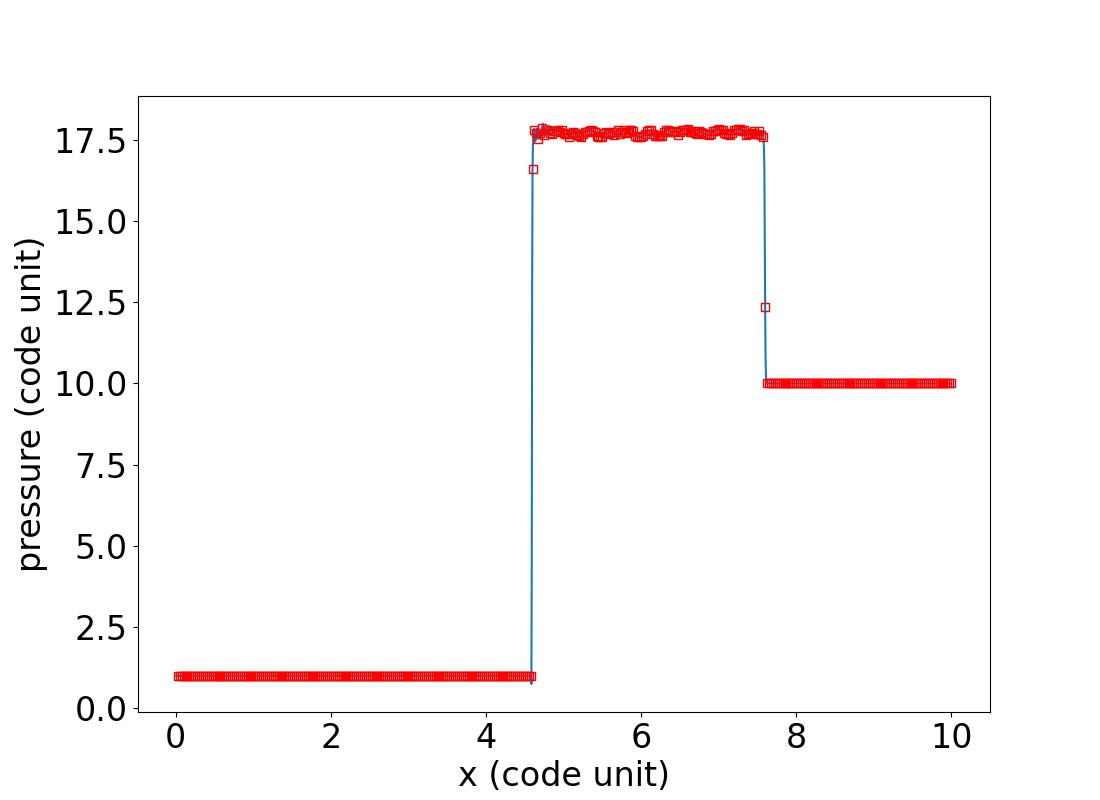}
\includegraphics*[width=5.7cm]{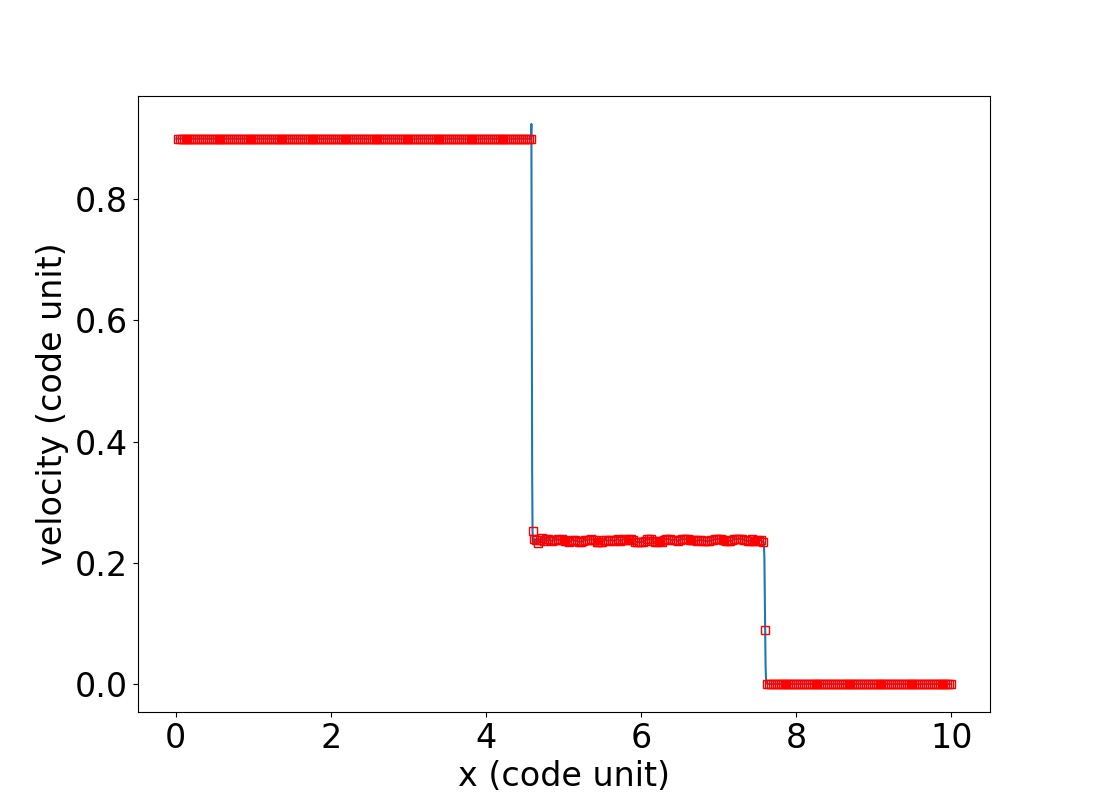}
\caption{Similar to Figure \ref{fig:Rosswog1_test} but from
the Riemann problem test case taken from the Test 3 in \cite{Marti1994}.}
\label{fig:Marti3_test}
\end{figure*}

To test the SR extension of our code, we perform a few standard
shock tube tests which aim at testing the fluid motion where
the velocity is close to the speed of light. The tests are described in Table \ref{table:SR_test}.

\begin{table}
\begin{center}
\caption{The input parameter for the special relativistic shock tube tests
performed in this work.}
\begin{tabular}{|c|c|c|c|c|c|c|c|c|}
\hline
Test & $\gamma$ & $x_{{\rm int}}$ & $\rho_L$ & $p_L$ & $v_L$ & $\rho_R$ & $p_R$ & $v_R$ \\ \hline
Test 1 & 5/3 & 5 & 10 & 40/3 & 0 & 1 & $10^{-6}$ & 0 \\
Test 2 & 5/3 & 5 & 1 & 10 & -0.6 & 10 & 20 & 0.5 \\  
Test 3 & 4/3 & 5 & 1 & 1 & 0.9 & 1 & 10 & 0 \\ \hline

\end{tabular}
\label{table:SR_test}
\end{center}

\end{table}

In Figure \ref{fig:Rosswog1_test} we plot the density, velocity 
and pressure of Test 1. 
The test is taken from the first test presented in \cite{Rosswog2009}.
This is the ``standard'' blast test which is frequently employed to 
test the shock-capturing ability of the Riemann solver in the 
special relativistic regime. The test features a high density and 
high pressure matter on the left and low density and low pressure 
matter on the right. This creates a shock propagating to the 
right with a velocity $\sim 0.7 c$. 

In Figure \ref{fig:Marti2_test} we plot similar to Figure \ref{fig:Rosswog1_test}
but for the test taken from the second test in \cite{Marti1994}.
This test features two flows moving away from each other, creating pressure
and density discontinuities. 

In Figure \ref{fig:Marti3_test} we plot similar to Figure \ref{fig:Rosswog1_test}
but for the test taken from the third test in \cite{Marti1994}.
This test approaches to a high velocity $v = 0.9 c$ on the left 
and a pressure discontinuity. 
The setting reproduces the multi-step structure and the results does not show an observable over- or undershooting across the discontinuities.


\bibliographystyle{aasjournal}
\pagestyle{plain}
\bibliography{biblio}

\end{document}